\documentclass[12pt,letterpaper]{article}
\usepackage{graphicx}
\usepackage{accents}

\def\title#1{{\bf\Large #1}}
\def\sec#1{\vspace{1.00\baselineskip} \noindent {\bf\large #1}
\vspace{0.25\baselineskip}}
\def\subsec#1{\vspace{0.50\baselineskip} \noindent {\bf #1}}

\def\al{\alpha}
\def\bt{\beta}
\def\ga{\gamma}

\def\de{\delta}

\def\ep{\epsilon}
\def\et{\eta}

\def\im{{\rm Im}}

\def\la{\lambda}

\def\om{\omega}

\def\pa{\parallel}
\def\pe{\perp}
\def\re{{\rm Re}}
\def\rh{\rho}

\def\pd{\partial}
\def\ph{\phi}

\def\si{\sigma}

\def\ta{\tau}
\def\th{\theta}

\def\vh{\vec{h}}
\def\vs{\vec{s}}
\def\vsi{\vec{\si}}
\def\ze{\zeta}

\def\d{\dagger}
\def\<{\langle}
\def\>{\rangle}

\def\tsum{{\textstyle\sum}}

\def\d{\dagger}
\def\tr{{\rm tr}}
\def\diag{{\rm diag}}

\def\vm{\vec{m}}
\def\vn{\vec{n}}

\def\vv{\vec{v}}
\def\vw{\vec{w}}

\def\pa{\parallel}
\def\pe{\perp}

\def\ba{\begin{eqnarray}}
\def\ea{\end{eqnarray}}
\def\be{\begin{equation}}
\def\ee{\end{equation}}

\begin{document}


\begin{center}

\vspace*{0.5\baselineskip}

\title{Exponentiation and decomposition formulas for \\ common operators 1: Classical applications}

\vspace{1.0\baselineskip}

C. J. McKinstrie \\

{\it\small Independent Photonics Consultant, Manalapan, NJ 07726, United States}

\vspace{0.50\baselineskip}

M. V. Kozlov

{\it\small Center for Preparatory Studies, Nazarbayev University, Astana 010000, Kazakhstan}

\vspace{0.50\baselineskip}

Abstract \\

\vspace{0.50\baselineskip}

\parbox[]{6.5in}{\small  In this tutorial, exponentiation and factorization (decomposition) formulas are derived and discussed for common matrix operators that arise in studies of classical dynamics, linear and nonlinear optics, and special relativity. 
To understand the physical properties of systems of common interest, one first needs to understand the mathematical properties of the symplectic group Sp(2), the special unitary groups SU(2) and SU(1,1), and the special orthogonal groups SO(3) and SO(1,2).
For these groups, every matrix can be written as the exponential of a generating matrix, which is a linear combination of three fundamental matrices (generators). For Sp(2), SU(1,1) and SO(1,2), every matrix also has a Schmidt decomposition, in which it is written as the product of three simpler matrices. The relations between the entries of the matrix, the generator coefficients and, where appropriate, the Schmidt-decomposition parameters are described in detail.
It is shown that Sp(2) is isomorphic to (has the same structure as) SU(1,1) and SO(1,2), and SU(2) is isomorphic to SO(3). Several examples of these isomorphisms (relations between Schmidt decompositions and product rules) are described, which illustrate their usefulness (complicated results can be anticipated or derived easily). This tutorial is written at a level that is suitable for senior undergraduate students and junior graduate students.}

\end{center}

\newpage

\sec{1. Introduction}

Group theory has many applications in the physical sciences \cite{ham89,tin03}. Although the mathematics of group theory is interesting in its own right, in our opinion, group theory is important because it facitilates the modeling, understanding and classification of physical processes. In this tutorial, we review the properties of several groups, which we have encountered in our studies of classical dynamics, linear, nonlinear and quantum optics, and special relativity.

The symplectic group Sp(2) arises in classical dynamics \cite{gol01,tay05} and geometrical (ray) optics \cite{buc93,bor99}.
The special unitary group SU(2) arises in optical beam splitting \cite{pra87,ouz89}, frequency conversion by three- and four-wave mixing \cite{mar08,boy20,wod85,ger01}, and polarization optics in Jones space \cite{gor00,chi18}.
The indefinite unitary group SU(1,1) arises in parametric amplification (one- and two-mode squeezing) by three- and four-wave mixing \cite{mar08,boy20,wod85,ger01}.
The special orthogonal group SO(3) governs rotations in three dimensions, and polarization optics in Stokes space \cite{gor00,chi18}.
The indefinite orthogonal groups SO(1,1), SO(1,2) and SO(1,3) govern Lorentz transformations in time, and one, two and three space dimensions, respectively \cite{jac99,lan13}. A discussion of SO(1,2) is included, because its complexity is intermediate between those of SO(1,1) and SO(1,3).

The matrices in the aforementioned groups are distinct: Sp(2) consists of real $2 \times 2$ matrices, SU(2) and SU(1,1) consist of complex $2 \times 2$ matrices, and SO(3) and SO(1,2) consist of real $3 \times 3$ matrices. 
The real  groups are defined by equations of the form $M^tSM = S$, where S is a real structure matrix. (Each group has a different structure matrix.) For the complex groups, $M^t$ is replaced by $M^\d$. By considering these equations, one finds that the members of each group are specified by three real parameters (three real coefficients, or three real or imaginary parts). Every matrix $M$ can be written as the exponential of a generating matrix $G$. The real generators are defined by equations of the form $G^tS + SG = 0$. For the complex groups, $G^t$ is replaced by $G^\d$. By considering these equations, one finds that every generating matrix can be written as the linear combination $G = G_1k_1 + G_2k_2 + G_3k_3$, where $G_i$ is a fundamental (basis) generator and $k_i$ is a real generator coefficient.
The matrices form (continuous) Lie groups, whereas the generating matrices form Lie algebras \cite{sat86,gil08}.

In most of the aforementioned applications, the generating matrices arise naturally as coefficient matrices in the matrix differential equations that govern the processes of interest \cite{bra83}. It is clear that the properties of the generators are impressed upon the solution matrices (which are also called Green or transfer matrices). In ray optics, one uses the laws of geometric optics (reflection and refraction) to derive transfer matrices for optical elements (such as lenses, mirrors and spaces), and it is these matrices that arise naturally \cite{ger94,pea15}. The transfer matrix for a composite system is the product of the constituent transfer matrices.

This tutorial is organized as follows: In Sec. 2, the matrix groups Sp(2), SU(2), SU(1,1), SO(3) and SO(1,2) are introduced. The canonical forms of these matrices and their generators are stated and discussed.
In Sec. 3, the Cayley--Hamilton (CH) theorem \cite{gil08,hor13} is used to exponentiate the $2 \times 2$ generators of Sp(2), SU(2) and SU(1,2). Exponentiation produces the canonical forms of these matrices, in which the matrix components are functions of the generator coefficients. In Sec. 4, the CH theorem is used to exponentiate the $3 \times 3$ generators of SO(3) and SO(1,2).

Every real matrix has the Schmidt decomposition $M = QDP^t$, where $D$ is diagonal, and $P$ and $Q$ are orthogonal \cite{hor13}. Likewise, every complex matrix has the decomposition $M = VDU^\d$, where $U$ and $V$ are unitary \cite{hor13,mck13}. In Sec. 5, Schmidt decompositions are derived for matrices in Sp(2), SU(1,1) and SO(1,2). Each matrix is specified by one dilation parameter ($\la$) and two angle parameters ($\th_1$ and $\th_2$). Similar (triple-product) decompositions are derived for matrices in SO(3). Each matrix, which corresponds to a rotation about an arbitrary axis, is specified by three angles, which correspond to rotations about the coordinate axes.

It is clear from the preceding discussion that every matrix of interest can be specified in terms of the matrix entries (components), the generator coefficients or the decomposition parameters (coefficients). Each of the three matrix representations has its advantages and disadvantages. Consequently, it is important to relate the components and two sets of coefficients, so one can use the representation that provides the most physical insight.

Groups whose generators satisfy the same commutation relations are isomorphic (have the same structure) \cite{kim83}. In Sec. 6, it is shown that Sp(2) is isomorphic to SU(1,1) and SO(1,2), and SU(2) is isomorphic to SO(3). For members of isomorphic groups, the relations between the decomposition and generator coefficients are equivalent. 

Group theory is taught regularly to students of some subfields of physics, but it is not necessarily taught to students of other subfields of physics, or engineering. When we decided to write this tutorial, our goal was to collect the minimum required knowledge of group theory in one place, and provide enough examples to demonstrate its power and usefulness.
We thought that it would be beneficial to discuss the five groups of interest together, to show their many similarities and few differences. The tutorial that resulted is longer than we anticipated (perhaps too long to be read at once). We suggest that Sec. 2 and the introductions to Secs. 3 -- 6 be read in their entirety, because the concepts described therein are universal. Subsequently, readers can pick examples related to the group(s) in which they are interested.
Several appendices are included, in which the transfer matrices of common physical systems are shown to be members of the groups discussed in the tutorial.

Discussions of the quantum operators $J_0$ and $J_\pm$, which arise in the theory of angular momentum \cite{sha94,gri18} and two-mode frequency conversion \cite{wod85,ger01}, and $K_0$ and $K_\pm$, which arise in the theory of one- and two-mode squeezing \cite{wod85,ger01}, were omitted from this tutorial. So also was a discussion of the differentiate-and-integrate method \cite{lou73,bar97}, which one can use to derive decomposition (disentanglement) formulas for exponentials of the $J$ and $K$ (and other) operators. We hope to rectify this shortcoming in a future tutorial, which will focus on quantum applications of group theory.

\newpage

\sec{2. Common matrix groups}

Consider a set of objects and a binary operation, which allows the objects to interact. The set is called a group if four conditions are satisfied \cite{taw95}.
First, the group is closed under the binary operation: If $A$ and $B$ are members of the set, then $C= B \circ A$ is also a member.
Second, the binary operation is associative: $C \circ (B \circ A) = (C \circ B) \circ A = C \circ B \circ A$.
Third, the set contains an identity element $I$, for which $I \circ A = A = A \circ I$. Fourth, every member $A$ has an inverse $A^{-1}$, for which $A^{-1} \circ A = I = A \circ A^{-1}$.
In this section, we describe the basic properties of common matrix groups and their generators.

\subsec{2.1. Symplectic matrices}

The special linear group SL(2) is the set of $2 \times 2$ real matrices with determinant 1, for which the binary operation is matrix multiplication.
A matrix $M$ is symplectic if it satisfies the equivalent equations
\be M^tJM = J, \ \ M^{-1} = J^tM^tJ, \label{2.1.1} \ee
where the structure matrix
\be J = \left[\begin{array}{cc} 0 & 1 \\ -1 & 0. \end{array}\right]. \label{2.1.2} \ee
Notice that $J^t = -J$, $J^2 = -I$ and $J^tJ = I$, so $J$ is orthogonal.

Symplectic matrices arise in Hamiltonian dynamics (App. A). 
Let $X = [x_1, x_2]^t$ and $Y = [y_1, y_2]^t$ be column vectors. Then the inner product $X^tY = x_1y_1 + x_2y_2$ and the cross-product $X^tJY = x_1y_2 - x_2y_1$. Transformations produced by symplectic matrices conserve the cross-product (phase-plane area), because $(MX)^tJ(MY) = X^t(M^tJM)Y = X^tJY$.

By substituting the ansatz
\be M = \left[\begin{array}{cc} \al & \bt \\ \ga & \de \end{array}\right] \label{2.1.3} \ee
in the first of Eqs. (\ref{2.1.1}), one finds that the symplectic condition is equivalent to the determinant condition $\al\de - \bt\ga = 1$. Hence, the symplectic group Sp(2) equals the special linear group SL(2). Both groups involve four real components and one real constraint, so they are three-parameter groups.

Every symplectic matrix can be written in the form
\be M(t) = \exp(Gt), \label{2.1.4} \ee
where $G$ is the generating matrix and $t$ is a parameter.
(In the context of matrix differential equations, $G$ is the coefficient matrix, $t$ is time and $M$ is the Green matrix \cite{bra83}.)
By substituting ansatz (\ref{2.1.4}) in the first of Eqs. (\ref{2.1.1}) and taking the limit $t \rightarrow 0$, one obtains the generator equation
\be G^tJ + JG = 0. \label{2.1.5} \ee
By writing $G$ in terms of its components, $a$, $b$, $c$ and $d$, one finds that $a + d = 0$, whereas $b$ and $c$ are arbitrary. Three matrices with these properties are
\be G_1 = \left[\begin{array}{cc} 1 & 0 \\ 0 & -1 \end{array}\right], \ \ 
G_2 = \left[\begin{array}{cc} 0 & 1 \\ 1 & 0 \end{array}\right], \ \ 
G_3 = \left[\begin{array}{cc} 0 & -1 \\ 1 & 0 \end{array}\right]. \label{2.1.6} \ee
The first matrix generates a dilation, the second generates a Lorentz boost and the third generates an active rotation ($G_3 = -J$).
Notice that Eq. (\ref{2.1.5}) is linear in $G$. Hence, if $G_1$ -- $G_3$ satisfy the equation, so also does the linear combination $G_1k_1 + G_2k_2 + G_3k_3$, where $k_1$ -- $k_3$ are arbitrary real numbers. The set of generating matrices is a vector space under addition \cite{taw91}, in which the matrices in Eq. (\ref{2.1.6}) play the role of basis vectors.
Notice also that these matrices have zero trace. It follows from the identity $\det(M) = \exp[\tr(G)]$ that $\det(M) = 1$ if and only if $\tr(G) = 0$. All the matrices considered in this article have unit determinant, so all their generators have zero trace. For symplectic matrices, the generators are subject to no additional constraints. 

The matrices in Eq. (\ref{2.1.6}) satisfy the commutation relations
\be [G_1, G_2] = -2G_3, \ \ [G_2, G_3] = 2G_1, \ \ [G_3, G_1] = 2G_2, \label{2.1.7} \ee
where the commutator $[x, y] = xy - yx$. Notice that the sign on the right side of the first of Eqs. (\ref{2.1.7}) is the opposite of the signs in the other equations. With the generators known, the matrix
\be M = \exp(G_1k_1 + G_2k_2 + G_3k_3), \label{2.1.8} \ee
where the coefficients $k_1$, $k_2$ and $k_3$ include the parameter $t$ (which is redundant). Equation (\ref{2.1.8}) also shows that the symplectic group is a three-parameter group.

\subsec{2.2. Unitary matrices}

A complex matrix is unitary if it satisfies the equivalent equations
\be M^\d M = I, \ \ M^{-1} = M^\d. \label{2.2.1} \ee
Unitary matrices conserve the inner product $X^\d Y$, because $(MX)^\d MY = X^\d(M^\d M)Y = X^\d Y$. They arise in models of beam splitting and frequency conversion (App. B). A unitary matrix can be written in the form
\be M = e^{i\ph} \left[\begin{array}{cc} \ta & \rh \\ -\rh^* & \ta^* \end{array}\right], \label{2.2.2} \ee
where $|\ta|^2 + |\rh|^2 = 1$. The group of such matrices is called the unitary group U(2). The special unitary group SU(2) is the subgroup of $U(2)$ whose members have determinant 1 ($\ph = 0$). The member matrices are specified by three real parameters: $|\ta|$, $\ph_\ta$ and $\ph_\rh$.

By substituting ansatz (\ref{2.1.4}) in the first of Eqs. (\ref{2.2.1}), one obtains the generator equation
\be G^\d + G = 0. \label{2.2.3} \ee
Hence, the generators are anti-Hermitian.
Three such matrices, with zero trace, are
\be G_1 = \left[\begin{array}{cc} i & 0 \\ 0 & -i \end{array}\right], \ \ 
G_2 = \left[\begin{array}{cc} 0 & i \\ i & 0 \end{array}\right], \ \ 
G_3 = \left[\begin{array}{cc} 0 & -1 \\ 1 & 0 \end{array}\right]. \label{2.2.4} \ee
The first matrix produces a differential phase shift, the second produces a beam-splitter-like transformation \cite{ouz89} and the third produces an active rotation. These matrices satisfy the commutation relations
\be [G_1, G_2] = 2G_3, \ \ [G_2, G_3] = 2G_1, \ \ [G_3, G_1] = 2G_2. \label{2.2.5} \ee
Notice that the signs on the right sides of Eqs. (\ref{2.2.5}) are all the same (positive).

If we had chosen the alternative ansatz $M = \exp(iH)$, we would have obtained the generator condition $ H^\d + H = 0$, from which it follows that $H$ is Hermitian. Three such matrices are the Pauli spin matrices
\be \si_1 = \left[\begin{array}{cc} 1 & 0 \\ 0 & -1 \end{array}\right], \ \ 
\si_2 = \left[\begin{array}{cc} 0 & 1 \\ 1 & 0 \end{array}\right], \ \ 
\si_3 = \left[\begin{array}{cc} 0 & -i \\ i & 0 \end{array}\right], \label{2.2.6} \ee
which satisfy the commutation relations
\be [\si_1, \si_2] = 2i\si_3, \ \ [\si_2, \si_3] = 2i\si_1, \ \ [\si_3, \si_1] = 2i\si_2. \label{2.2.7} \ee
Once again, the signs on the right sides of Eqs. (\ref{2.2.7}) are all the same. Notice that $\si_1 = G_1/i$ and $\si_2 = G_2/i$, but $\si_3 = iG_3$.

Now define the metric matrix $S = \diag(1, -1)$ and the generalized inner product $X^\d SY = x_1^*y_1 - x_2^*y_2$. The metric is termed indefinite, because the norm $X^\d SX$ can be negative. A complex matrix is indefinite unitary if it satisfies the equivalent equations
\be M^\d S M = S, \ \ M^{-1} = SM^\d S. \label{2.2.11} \ee
Indefinite unitary matrices conserve the generalized inner product, because $(MX)^\d S(MY) = X^\d(M^\d SM)Y = X^\d SY$. They arise in models of parametric amplification (one and two-mode squeezing) by three- and four-wave mixing (App. C).
An indefinite unitary matrix can be written in the form
\be M = e^{i\ph} \left[\begin{array}{cc} \mu & \nu \\ \nu^* & \mu^* \end{array}\right], \label{2.2.12} \ee
where $|\mu|^2 - |\nu|^2 = 1$. The set of such matrices is called the indefinite unitary group $U(1,1)$. The subgroup of U(1,1) whose members have  determinant 1 ($\ph = 0$) is called the special indefinite unitary group SU(1,1). The member matrices are specified by three real parameters: $|\mu|$, $\ph_\mu$ and~$\ph_\nu$.

By substituting ansatz (\ref{2.1.4}) in the first of Eqs. (\ref{2.2.11}), one obtains the generator equation
\be G^\d S + SG = 0. \label{2.2.13} \ee
The first term in Eq. (\ref{2.2.13}) is $(SG)^\d$, so $SG = iH$, or $G = iSH$. The generator is proportional to the product of a Hermitian matrix and the metric matrix.
By writing $G$ in terms of its components $a$, $b$, $c$ and $d$, one finds that $a$ and $d$ are imaginary (or zero), whereas $c = b^*$. Three matrices with these properties are
\be 
G_1 = \left[\begin{array}{cc} 0 & 1 \\ 1 & 0 \end{array}\right], \ \ 
G_2 = \left[\begin{array}{cc} 0 & i \\ -i & 0 \end{array}\right], \ \
G_3 = \left[\begin{array}{cc} i & 0 \\ 0 & -i \end{array}\right]. \label{2.2.14} \ee
These matrices satisfy the commutation relations
\be [G_1, G_2] = -2G_3, \ \ [G_2, G_3] = 2G_1, \ \ [G_3, G_1] = 2G_2, \label{2.2.15} \ee
Notice that relations (\ref{2.2.15}) are identical to relations (\ref{2.1.7}).

\subsec{2.3. Orthogonal matrices}

A real $3 \times 3$ matrix is orthogonal if it satisfies the equivalent equations
\be M^tM = I, \ \ M^{-1} = M^t. \label{2.3.1} \ee
It follows from the inverse condition that the columns $M$ are orthonormal vectors, as are the rows.
Orthogonal matrices conserve the inner product $X^tY$, because $(MX)^t(MY) = X^tM^tMY = X^tY$. They arise in three-dimensional rotation, in which context $\det(M) = 1$ and $X = [x, y, z]^t$.
A rotation matrix can be written in the form
\be M = \left[\begin{array}{ccc} c + n_1^2d & -n_3s + n_1n_2d & n_2s + n_1n_3d \\
n_3s + n_2n_1d & c + n_2^2d & -n_1s + n_2n_3d \\
-n_2s + n_3n_1d & n_1s + n_3n_2d & c + n_3^2d \end{array}\right], \label{2.3.2} \ee
where $(n_1, n_2, n_3)$ is the unit vector that defines the rotation axis, $c = \cos\th$, $s = \sin\th$, $d = 1 - c$ and $\th$ is the rotation angle (Sec. 4 and App. D). The group of such matrices is called the special orthogonal group SO(3). Its member matrices are specified by three real parameters: $\th$ and the two polar angles that specify the direction of the rotation axis.

By substituting the ansatz (\ref{2.1.4}) in the first of Eqs. (\ref{2.3.1}) and taking the limit as $t \rightarrow 0$, one obtains the generator equation
\be G^t + G = 0. \label{2.3.3} \ee
Hence, $G$ is anti-symmetric. Three such matrices are
\be G_1 = \left[\begin{array}{ccc} 0 & 0 & 0 \\ 0 & 0 & -1 \\ 0 & 1 & 0 \end{array}\right], \ \ 
G_2 = \left[\begin{array}{ccc} 0 & 0 & 1 \\ 0 & 0 & 0 \\ -1 & 0 & 0 \end{array}\right], \ \ 
G_3 = \left[\begin{array}{ccc} 0 & -1 & 0 \\ 1 & 0 & 0 \\ 0 & 0 & 0 \end{array}\right]. \label{2.3.4} \ee
These matrices produce active rotations about the $x$, $y$ and $z$ axes, respectively.
They satisfy the commutation relations
\be [G_1, G_2] = G_3, \ \ [G_2, G_3] = G_1, \ \ [G_3, G_1] = G_2, \label{2.3.5} \ee
Relations (\ref{2.3.5}) for SO(3) are equivalent to relations (\ref{2.2.5}) for SU(2), which one can verify by dividing the first set of generators by 2 or multiplying the second set by 2.

Now define the metric matrix $S = \diag(1, -1, -1)$ and the generalized inner product $X^tSY = x_1y_1 - x_2y_2 - x_3y_3$. The metric is termed indefinite, because the norm $X^tSX$ can be negative. A matrix is indefinite orthogonal if it satisfies the equivalent equations
\be M^tSM = S, \ \ M^{-1} = SM^tS. \label{2.3.11} \ee
Indefinite orthogonal matrices conserve the generalized inner product, because $(MX)^tS(MY)$ $= X^t(M^tSM)Y = X^tSY$. They arise in special relativity as Lorentz transformations in time and two space dimensions, in which context $\det(M) = 1$ and $X = [t, x, y]^t$. The columns of $M$ are the images of the vectors $[1, 0, 0]^t$, $[0, 1, 0]^t$ and $[0, 0, 1]^t$, which have generalized norms of 1, $-1$ and $-1$, respectively. In particular, the first column is a dimensionless energy--momentum vector. Similar remarks can be made about the columns of $M^t$ (rows of $M$). A transformation matrix can be written in the form
\be M = \left[\begin{array}{ccc} \ga & uc_1 & us_1 \\ uc_2 & c_{21} + \de c_2c_1 & -s_{21} + \de c_2s_1 \\ us_2 & s_{21} + \de s_2c_1 & c_{21} + \de s_2s_1 \end{array}\right], \label{2.3.12} \ee
where $\ga$ is the (dimensionless) energy, $u = (\ga^2 - 1)^{1/2}$ is the momentum, $c_i = \cos(\th_i)$, $s_i = \sin(\th_i)$ and $\de = \ga - 1$ (Sec. 4 and App. E). The angles $\th_1$ and $\th_2$ specify the directions of the momentum vectors, and the difference angle $\th_{21} = \th_2 - \th_1$. The group of such matrices is called the special indefinite orthogonal group SO(1,2), or the reduced Lorentz group. Its member matrices are specified by three real parameters: $\ga$, $\th_1$ and $\th_2$.

By substuting ansatz (\ref{2.1.4}) in the first of Eqs. (\ref{2.3.11}), one obtains the generator equation
\be G^tS + SG = 0. \label{2.3.13} \ee
The first term in Eq. (\ref{2.3.13}) is $(SG)^t$, from which it follows that $G = SA$, where $A$ is anti-symmetric. Three matrices with this property are
\be G_1 = \left[\begin{array}{ccc} 0 & 1 & 0 \\ 1 & 0 & 0 \\ 0 & 0 & 0 \end{array}\right], \ \ 
G_2 = \left[\begin{array}{ccc} 0 & 0 & 1 \\ 0 & 0 & 0 \\ 1 & 0 & 0 \end{array}\right], \ \ 
G_3 = \left[\begin{array}{ccc} 0 & 0 & 0 \\ 0 & 0 & -1 \\ 0 & 1 & 0 \end{array}\right]. \label{2.3.14} \ee
The first matrix generates a boost in the $x$ direction, the second generates a boost in the $y$ direction and the third generates a rotation about the $t$ axis (in the $xy$ plane).
These matrices satisfy the commutation relations
\be [G_1, G_2] = -G_3, \ \ [G_2, G_3] = G_1, \ \ [G_3, G_1] = G_2. \label{2.3.15} \ee
Relations (\ref{2.3.15}) for SO(1,2) are equivalent to relations (\ref{2.1.7}) for Sp(2) and relations (\ref{2.2.15}) for SU(1,1), which one can verify by dividing the first and second sets of generators by 2 or multiplying the third set by 2.

The results of Secs. 2.1 -- 2.3 are summarized in Tab. 1. The commutation relations for Sp(2), SU(1,1) and SO(1,2) are equivalent, as are the relations for SU(2) and SO(3).
\begin{table}[h!]
\centering
\begin{tabular}{|c|c|c|c|}
\hline
Group & Matrix & Generator & Parameters \\
\hline
Sp(2) & $M^tJM = J$ & $G^tJ + JG = 0$ & 3 \\
\hline
SU(2) & $M^\d M = I$ & $G^\d + G = 0$ & 3 \\
\hline
SU(1,1) & $M^\d SM = S$ & $G^\d S + SG = 0$ & 3 \\
\hline
SO(3) & $M^tM = I$ & $G^t + G = 0$ & 3 \\
\hline
SO(1,2) & $M^tSM = S$ & $G^tS + SG = 0$ & 3 \\
\hline
\end{tabular}
\vspace*{0.05in}
\caption{Defining properties of the famous five groups. Sp(2), SO(3) and SO(1,2) are real, whereas SU(2) and SU(1,1) are complex. Each group involves three real parameters. The structure matrices $J$ and $S$ are defined in the text. }
\end{table}

\vspace*{3in}

\newpage

\sec{3. Exponentiation of the $2 \times 2$ generators}

One can exponentiate a $2 \times 2$ matrix by using the Cayley--Hamilton (CH) theorem \cite{gil08}. Let $\la_1$ and $\la_2$ be the eigenvalues of $G$. Then $G$ satisfies the characteristic equation
\be G^2 - (\la_1 + \la_2)G + \la_1\la_2 I = 0, \label{3.1.3} \ee
from which it follows that
\be G^2 = -\la_1\la_2 I + (\la_1 + \la_2)G. \label{3.1.4} \ee
Hence, $\exp(G) = aI + bG$, where $a$ and $b$ are functions of the eigenvalues.

\subsec{3.1 Symplectic matrices}

For Sp(2,R), the generators are
\be G_1 = \left[\begin{array}{cc} 1 & 0 \\ 0 & -1 \end{array}\right], \ \ 
G_2 = \left[\begin{array}{cc} 0 & 1 \\ 1 & 0 \end{array}\right], \ \ 
G_3 = \left[\begin{array}{cc} 0 & -1 \\ 1 & 0 \end{array}\right]. \label{3.1.1} \ee
Let $G = G_1k_1 + G_2k_2 + G_3k_3$. Then, written explicitly, the generating matrix
\be G = \left[\begin{array}{cc} k_1 & k_2 - k_3 \\ k_2 + k_3 & -k_1 \end{array}\right]. \label{3.1.2} \ee
It is easy to verify that the eigenvalues of $G$ are $\pm k$, where $k = (k_1^2 + k_2^2 - k_3^2)^{1/2}$. According to the CH theorem, $G^2 = k^2I$. Hence,
\ba \exp(G) &= &I + G + k^2I/2 + k^2G/3! \ \dots \nonumber \\
&= &I\cosh(k) + G\sinh(k)/k. \label{3.1.3} \ea
Written explicitly, the exponentiated matrix
\be M = \left[\begin{array}{cc} C + n_1S & (n_2 - n_3)S \\ (n_2 + n_3)S & C - n_1S \end{array}\right], \label{3.1.4} \ee
where $C = \cosh(k)$, $S = \sinh(k)$ and $n_i = k_i/k$. It is easy to verify that $\det(M) = 1$.

It follows from Eq. (\ref{3.1.3}) that the inverse matrix
\be \exp(-G) = I\cosh(k) - G\sinh(k)/k. \label{3.1.5} \ee
By combining Eqs. (\ref{3.1.3}) and (\ref{3.1.5}), one finds that
\ba e^{-G}e^G &= & (IC - GS/k)(IC + GS/k) \nonumber \\
&= &IC^2 - G^2S^2/k^2 \nonumber \\
&= &I, \label{3.1.6} \ea
because $G^2 = k^2I$ and $C^2 - S^2 = 1$. Written explicitly, the inverse matrix
\be M^{-1} = \left[\begin{array}{cc} C - n_1S & -(n_2 - n_3)S \\ -(n_2 + n_3)S & C + n_1S \end{array}\right], \label{3.1.7} \ee
By comparing formulas (\ref{3.1.4}) and (\ref{3.1.7}), one finds that the matrix and its inverse are related by the standard rule for $2 \times 2$ matrices (as they should be). Formula (\ref{3.1.7}) is also consistent with the second of Eqs. (\ref{2.1.1}).

The derivation of formula (\ref{3.1.4}) was based on the assumption that $k_1^2 + k_2^2 - k_3^2 > 0$. In the opposite case, $k \rightarrow ik = i(k_3^2 - k_1^2 - k_2^3)^{1/2}$, $\cosh(k) \rightarrow \cos(k)$ and $\sinh(k)/k \rightarrow \sin(k)/k$.
With these changes, formulas (\ref{3.1.4}) and (\ref{3.1.7}) remain valid. ($G^2 \rightarrow -k^2I$, because the definition of $k$ changes.)

\subsec{3.2 Unitary matrices}

For SU(2), the generators are
\be G_1 = \left[\begin{array}{cc} i & 0 \\ 0 & -i \end{array}\right], \ \ 
G_2 = \left[\begin{array}{cc} 0 & i \\ i & 0 \end{array}\right], \ \ 
G_3 = \left[\begin{array}{cc} 0 & -1 \\ 1 & 0 \end{array}\right], \label{3.2.1} \ee
and the generating matrix
\be G = \left[\begin{array}{cc} ik_1 & ik_2 - k_3 \\ ik_2 + k_3 & -ik_1 \end{array}\right]. \label{3.2.2} \ee
It is easy to verify that the eigenvalues of $G$ are $\pm ik$, where $k = (k_1^2 + k_2^2 + k_3^2)^{1/2}$. According to the CH theorem, $G^2 = -k^2I$. Hence,
\ba \exp(G) &= &I + G - k^2I/2 - k^2G/3! \ \dots \nonumber \\
&= &I\cos(k) + G\sin(k)/k. \label{3.2.3} \ea
Written explicitly, the exponentiated matrix
\be M = \left[\begin{array}{cc} c + in_1s & in_2s - n_3s \\ in_2s + n_3s & c - in_1s \end{array}\right]
= \left[\begin{array}{cc} c + i\de s/k & i\ga s/k \\ i\ga^*s/k & c - i\de s/k \end{array}\right], \label{3.2.4} \ee
where $c = \cos(k)$, $s = \sin(k)$, $n_i = k_i/k$, $\de = k_1$ and $\ga = k_2 + ik_3$ (App. B). Matrix (\ref{3.2.4}) has the correct form for a unitary matrix ($m_{21} = -m_{12}^*$ and $m_{22} = m_{11}^*$). Changing the sign of $G$ is equivalent to changing the signs of $n_i$. Hence, the inverse matrix
\be M^{-1} = \left[\begin{array}{cc} c - in_1s & -in_2s + n_3s \\ -in_2s - n_3s & c + in_1s \end{array}\right]. \label{3.2.5} \ee
By comparing formulas (\ref{3.2.4}) and (\ref{3.2.5}), one finds that $M^{-1} = M^\d$ (as it should do).
For SU(2), $k^2 \ge 0$, so there is no complementary case to consider.

\subsec{3.3 Indefinite unitary matrices}

For SU(1,1), the generators are
\be G_1 = \left[\begin{array}{cc} 0 & 1 \\ 1 & 0 \end{array}\right], \ \ 
G_2 = \left[\begin{array}{cc} 0 & i \\ -i & 0 \end{array}\right], \ \ 
G_3 = \left[\begin{array}{cc} i & 0 \\ 0 & -i \end{array}\right], \label{3.3.1} \ee
and the generating matrix
\be G = \left[\begin{array}{cc} ik_3 & k_1 + ik_2 \\ k_1 - ik_2 & -ik_3 \end{array}\right]. \label{3.3.2} \ee
It is easy to verify that the eigenvalues of $G$ are $\pm k$, where $k = (k_1^2 + k_2^2 - k_3^2)^{1/2}$. According to the CH theorem, $G^2 = k^2I$. Hence,
\ba \exp(G) &= &I + G + k^2I/2 + k^2G/3! \ \dots \nonumber \\
&= &I\cosh(k) + G\sinh(k)/k. \label{3.3.3} \ea
Written explicitly, the exponentiated matrix
\be M = \left[\begin{array}{cc} C + in_3S & (n_1 + in_2)S \\ (n_1  - in_2)S & C - in_3S \end{array}\right]
= \left[\begin{array}{cc} C + i\de S/k & i\ga S/k \\ -i\ga^*S/k & C - i\de S/k \end{array}\right], \label{3.3.4} \ee
where $C = \cosh(k)$, $S = \sinh(k)$, $n_i = k_i/k$, $\de = k_3$ and $i\ga = k_1 + ik_2$ (App. C). Matrix (\ref{3.3.4}) has the correct form for an indefinite unitary matrix ($m_{21} = m_{12}^*$ and $m_{22} = m_{11}^*$). Changing the sign of $G$ is equivalent to changing the signs of $n_i$. Hence, the inverse matrix
\be M^{-1} = \left[\begin{array}{cc} C - in_3S & -(n_1 + in_2)S \\ -(n_1 - in_2)S & C + in_3S \end{array}\right]. \label{3.3.5} \ee
Formula (\ref{3.3.5}) is consistent with the second of Eqs. (\ref{2.2.11}).

The derivation of formula (\ref{3.3.4}) was based on the assumption that $k_1^2 + k_2^2 - k_3^2 > 0$. In the opposite case, $k \rightarrow ik = i(k_3^2 - k_1^2 - k_2^3)^{1/2}$, $\cosh(k) \rightarrow \cos(k)$ and $\sinh(k)/k \rightarrow \sin(k)/k$.
With these changes, formulas (\ref{3.3.4}) and (\ref{3.3.5}) remain valid. ($G^2 \rightarrow -k^2I$, because the definition of $k$ changes.)

In this section, we determined how the matrices in Sp(2), SU(2) and SU(1,1) depend on the generator coefficients. It is also worthwhile to consider the inverse (dial-up) problem: If a matrix is specified, can one determine the coefficients required to produce it?
First, let $\al$, $\bt$, $\ga$ and $\de$ be the components of the symplectic matrix (\ref{2.1.3}). Then, it follows from Eq. (\ref{3.1.6}) that $C = (\al + \de)/2$, which determines $S = (C^2 - 1)^{1/2}$ and $k = \log(C + S)$. In turn, $n_1 = (\al - \de)/2S$, $n_2 = (\bt + \ga)/2S$ and $n_3 = (\ga - \bt)/2S$, where $n_i = k_i/k$.
Second, let $\ta$ and $\rh$ be the components of the unitary matrix (\ref{2.2.2}). Then it follows from Eq. (\ref{3.2.4}) that $c = \re(\ta)$, which determines $s = (1 - c^2)^{1/2}$ and $k = \log(c + is)/i$. In turn, $n_1 = \im(\ta)/s$, $n_2 = \im(\rh)/s$ and $n_3 = -\re(\rh)/s$.
Third, let $\mu$ and $\nu$ be the components of the indefinite unitary matrix (\ref{2.2.12}). Then it follows from Eq. (\ref{3.3.4}) that $C = \re(\mu)$, which determines $S = (C^2 - 1)^{1/2}$ and $k = \log(C + S)$. In turn, $n_1 = \im(\mu)/S$, $n_2 = \im(\nu)/S$ and $n_3 = -\re(\nu)/S$. For the $2 \times 2$ matrices, the inverse formulas are simple.

\newpage

\sec{4. Exponentiation of the $3 \times 3$ operators}

One can also exponentiate a $3 \times 3$ matrix by using the CH theorem. Let $\la_1$, $\la_2$ and $\la_3$ be the eigenvalues of $G$. Then $G$ satisfies the characteristic equation
\be G^3 - (\la_1 + \la_2 + \la_3)G^2 + (\la_1\la_2 + \la_2\la_3 + \la_3\la_1)G - \la_1\la_2\la_3I = 0, \label{4.1.3} \ee
from which it follows that
\be G^3 = \la_1\la_2\la_3I - (\la_1\la_2 + \la_2\la_3 + \la_3\la_1)G + (\la_1 + \la_2 + \la_3)G^2. \label{4.1.4} \ee
Hence, $\exp(G) = aI + bG + cG^2$, where $a$, $b$ and $c$ are functions of the eigenvalues.

\subsec{4.1 Orthogonal matrices}

For SO(3), the generators are
\be G_1 = \left[\begin{array}{ccc} 0 & 0 & 0 \\ 0 & 0 & -1 \\ 0 & 1 & 0 \end{array}\right], \ \ 
G_2 = \left[\begin{array}{ccc} 0 & 0 & 1 \\ 0 & 0 & 0 \\ -1 & 0 & 0 \end{array}\right], \ \ 
G_3 = \left[\begin{array}{ccc} 0 & -1 & 0 \\ 1 & 0 & 0 \\ 0 & 0 & 0 \end{array}\right]. \label{4.1.1} \ee
Let $G = G_1k_1 + G_2k_2 + G_3k_3$. Then, written explicitly, the generating matrix
\be G = \left[\begin{array}{ccc} 0 & -k_3 & k_2 \\ k_3 & 0 & -k_1 \\ -k_2 & k_1 & 0 \end{array}\right]. \label{4.1.2} \ee
It is easy to verify that $G$ has the eigenvalues 0 and $\pm ik$, where $k = (k_1^2 + k_2^2 + k_3^2)^{1/2}$. Hence, the sum and product of the eigenvalues are zero, and $G^3 = -k^2G$. The exponential
\ba \exp(G) &= & I + G + G^2/2 - Gk^2/3! - G^2k^2/4! + Gk^4/5! + G^2k^4/6! \dots \nonumber \\
&= &I + G(1 - k^2/3! + k^4/5! \dots) + G^2(1/2 - k^2/4! + k^4/6! \dots) \nonumber \\
&= &I + G\sin(k)/k + G^2[1 - \cos(k)]/k^2. \label{4.1.5} \ea
The squared matrix
\be G^2 = \left[\begin{array}{ccc} -k_2^2 - k_3^2 & k_1k_2 & k_1k_3 \\ k_2k_1 & -k_3^2 - k_1^2 & k_2k_3 \\ k_3k_1 & k_3k_2 & -k_1^2 - k_2^2 \end{array}\right]. \label{4.1.6} \ee
By combining the preceding results, one finds that
\ba \exp(G) &= &\left[\begin{array}{ccc} 1 - (n_2^2 + n_3^2)d & -n_3s + n_1n_2d & n_2s + n_1n_3d \\ n_3s + n_2n_1d & 1 - (n_3^2 + n_1^2)d & -n_1s + n_2n_3d \\ -n_2s + n_3n_1d & n_1s + n_3n_2d & 1 - (n_1^2 + n_2^2)d \end{array}\right] \nonumber \\
&= &\left[\begin{array}{ccc} c + n_1^2d & -n_3s + n_1n_2d & n_2s + n_1n_3d \\ n_3s + n_2n_1d & c + n_2^2d & -n_1s + n_2n_3d \\ -n_2s + n_3n_1d & n_1s + n_3n_2d & c + n_3^2d \end{array}\right], \label{4.1.7} \ea
where $c = \cos(k)$, $s = \sin(k)$, $d = 1 - c$ and $n_i = k_i/k$. Equation (\ref{4.1.7}) is identical to Eq. (\ref{2.3.2}): The generating-matrix method produces the canonical (and simplest) form of the rotation matrix naturally. The axis-angle parameters are related to the generator coefficients by the identities $\th = k$ and $n_i = k_i/k$, so there is no inverse problem to solve.

It follows from Eq. (\ref{4.1.5}) that the inverse matrix
\be \exp(-G) = I - G\sin(k)/k + G^2[1 - \cos(k)]/k^2. \label{4.1.9} \ee
By combining Eqs. (\ref{4.1.5}) and (\ref{4.1.9}), one finds that
\ba e^{-G}e^G &= &(I - Gs/k + G^2d/k^2)(I + Gs/k + G^2d/k^2) \nonumber \\
&= &(I + G^2d/k^2)^2 - G^2s^2/k^2 \nonumber \\
&= &I + 2G^2d/k^2 + G^4d^2/k^4 - G^2s^2/k^2. \label{4.1.10} \ea
By using the CH identity $G^3 = -k^2G$, one can rewrite the right side as the sum of $I$ and a term proportional to $G^2/k^2$. The coefficient is
\ba 2d - d^2 - s^2 = 2(1 - c)  - (1 - 2c + c^2) - s^2 = 0. \label{4.1.11} \ea
Written explicitly, the inverse matrix
\be M^{-1} = \left[\begin{array}{ccc} c + n_1^2d & n_3s + n_1n_2d &- n_2s + n_1n_3d \\ -n_3s + n_2n_1d & c + n_2^2d & n_1s + n_2n_3d \\ n_2s + n_3n_1d & -n_1s + n_3n_2d & c + n_3^2d \end{array}\right]. \label{4.1.12} \ee
For rotations, changing the sign of $G$ is equivalent to changing the sign of $\sin\th$, or the signs of $n_i$. This result is sensible. The inverse of a rotation of angle $\th$ about the axis $\vn$ is a rotation of angle $ \th$ about the same axis or a rotation of angle $\th$ about the anti-parallel axis $ \vn$. Notice that $M^{-1} = M^t$ (as it should do).

\subsec{4.2 Indefinite orthogonal matrices}

For SO(1,2), the generators are
\be G_1 = \left[\begin{array}{ccc} 0 & 1 & 0 \\ 1 & 0 & 0 \\ 0 & 0 & 0 \end{array}\right], \ \ 
G_2 = \left[\begin{array}{ccc} 0 & 0 & 1 \\ 0 & 0 & 0 \\ 1 & 0 & 0 \end{array}\right], \ \ 
G_3 = \left[\begin{array}{ccc} 0 & 0 & 0 \\ 0 & 0 & -1 \\ 0 & 1 & 0 \end{array}\right]. \label{4.2.1} \ee
The generating matrix and its square are
\be G = \left[\begin{array}{ccc} 0 & k_1 & k_2 \\ k_1 & 0 & -k_3 \\ k_2 & k_3 & 0 \end{array}\right], \ \ 
G^2 = \left[\begin{array}{ccc} k_1^2 + k_2^2 & k_2k_3 & -k_1k_3 \\ -k_3k_2 & k_1^2 - k_3^2 & k_1k_2 \\ k_3k_1 & k_2k_1 & k_2^2 - k_3^2 \end{array}\right]. \label{4.2.2} \ee
It is easy to verify that $G$ has the eigenvalues 0 and $\pm k$, where $k = (k_1^2 + k_2^2 - k_3^2)^{1/2}$. It follows from Eq. (\ref{4.1.4}) that $G^3 = k^2G$. Hence, the exponential
\ba \exp(G) &= & I + G + G^2/2 + Gk^2/3! + G^2k^2/4! + Gk^4/5! + G^2k^4/6! \dots \nonumber \\
&= &I + G(1 + k^2/3! + k^4/5! \dots) + G^2(1/2 + k^2/4! + k^4/6! \dots) \nonumber \\
&= &I + G\sinh(k)/k + G^2[\cosh(k) - 1)]/k^2. \label{4.2.3} \ea
By combining Eqs. (\ref{4.2.2}) and (\ref{4.2.3}), one obtains the exponentiated matrix
\ba M &= &\left[\begin{array}{ccc} 1 + (n_1^2 + n_2^2)D & n_1S + n_2n_3D & n_2S - n_1n_3D \\
n_1S - n_3n_2D & 1 + (n_1^2 - n_3^2)D & -n_3S + n_1n_2D \\
n_2S + n_3n_1D & n_3S + n_2n_1D & 1 + (n_2^2 - n_3^2)D \end{array}\right] \nonumber \\
&= &\left[\begin{array}{ccc} C + n_3^2D & n_1S + n_2n_3D & n_2S - n_1n_3D \\
n_1S - n_3n_2D & C - n_2^2D & -n_3S + n_1n_2D \\
n_2S + n_3n_1D & n_3S + n_2n_1D & C - n_1^2D \end{array}\right], \label{4.2.4} \ea
where $C = \cosh(k)$, $D = C - 1$, $S = \sinh(k)$ and $n_i = k_i/k$.

First, consider the special case in which $k_3 = 0$ and $k = (k_1^2 + k_2^2)^{1/2}$. Then
\be M = \left[\begin{array}{ccc} C & n_1S & n_2S \\ n_1S & 1 + n_1^2D & n_1n_2D \\ n_2S & n_1n_2D & 1 + n_2^2D \end{array}\right]. \label{4.2.5} \ee
Matrix (\ref{4.2.5}) decribes a Lorentz transformation (boost) with energy $\ga = C$ and momentum $u = S$ in the direction $(n_1, n_2) = (\cos\th, \sin\th)$ \cite{mck25b}. Conversely, if $\ga$ and $\th$ are specified, then $k = \log(\ga + u)$, $k_1 = k\cos\th$ and $k_2 = k\sin\th$. Notice that $\tr(M) = 2C + 1$ is positive.

Second, consider the complementary case in which $k_1 = k_2 = 0$ and $k = (-k_3^2)^{1/2} = ik_3$. Then
\be M = \left[\begin{array}{ccc} 1 & 0 & 0 \\ 0 & c_3  & -s_3 \\ 0 & s_3 & c_3 \end{array}\right], \label{4.2.6} \ee
where $c_3 = \cos(k_3)$ and $s_3 = \sin(k_3)$. Matrix (\ref{4.2.6}) describes a rotation through the angle $\th = k_3$, so there is no inverse problem to solve. Notice that $\tr(M) = 2c_3 + 1$ can be negative.

Third, consider the general case in which $n_1$, $n_2$, and $n_3 \neq 0$. By using the identities $n_1^2 + n_2^2 - n_3^2 = 1$ and $S^2 = D(D + 2)$, one can show that $m_{11}^2 - m_{12}^2 - m_{13}^2 = 1$ and $m_{11}^2 - m_{21}^2 - m_{31}^2 = 1$. Hence, the first row of matrix (\ref{4.2.4}) can be written in the form $[\ga, uc_1, us_1]$ and the first column can be written in the form $[\ga, uc_2, us_2]^t$, as stated in Eq. (\ref{2.3.12}). The energy and momentum are
\be \ga = 1 + (n_1^2 + n_2^2)D, \ \ u =  [(n_1^2 + n_2^2)(S^2 + n_3^2D^2)]^{1/2}, \label{4.2.7} \ee
respectively, and the angles are specified implicitly by the equations
\be \tan\th_1 = {n_2S - n_1n_3D \over n_1S + n_2n_3D}, \ \ 
\tan\th_2 = {n_2S + n_1n_3D \over n_1S - n_2n_3D}. \label{4.2.8} \ee
Notice that the only difference between these formulas is the sign of $n_3$.
It follows from Eqs. (\ref{4.2.8}), and the trigonometric identities $c = 1/(1 + t^2)^{1/2}$ and $s = t/(1 + t^2)^{1/2}$, that
\ba c_1 = {n_1S + n_2n_3D \over [(n_1^2 + n_2^2)(S^2 + n_3^2D^2)]^{1/2}}, \ \
s_1 = {n_2S - n_1n_3D \over [(n_1^2 + n_2^2)(S^2 + n_3^2D^2)]^{1/2}}, \label{4.2.9} \\
c_2 = {n_1S - n_2n_3D \over [(n_1^2 + n_2^2)(S^2 + n_3^2D^2)]^{1/2}}, \ \
s_2 = {n_2S + n_1n_3D \over [(n_1^2 + n_2^2)(S^2 + n_3^2D^2)]^{1/2}}. \label{4.2.10} \ea
Notice that the denominator in these formulas is $u$, so the numerators are $uc_i$ and $us_i$.

It remains to be shown that the formulas for the components of the lower-right block in Eqs. (\ref{2.3.12}) and (\ref{4.2.4}) are equivalent. (This proof is also provided in \cite{mck25b}.)
By combining Eqs. (\ref{4.2.8}), one finds that
\be \tan(\th_{21}) = {2n_3DS \over S^2 - n_3^2D^2}, \label{4.2.11} \ee
from which it follows that
\be c_{21} = {S^2 - n_3^2D^2 \over S^2 + n_3^2D^2}, \ \ 
s_{21} = {2n_3DS \over S^2 + n_3^2D^2}. \label{4.2.12} \ee
For the element $m_{22}$,
\ba c_{21} + \de c_2c_1 &= &{S^2 - n_3^2D^2 \over S^2 + n_3^2D^2}
+ (n_1^2 + n_2^2)D {(n_1S - n_2n_3D)(n_1S + n_2n_3D) \over [(n_1^2 + n_2^2)(S^2 + n_3^2D^2)]} \nonumber \\
&= &{S^2 - n_3^2D^2 \over S^2 + n_3^2D^2} + {(n_1^2S^2 - n_2^2n_3^2D^2)D \over S^2 + n_3^2D^2}. \label{4.2.13} \ea
The numerator in Eq. (\ref{4.2.13}) is
\ba &&S^2 - n_3^2D^2 + (n_1^2 + n_2^2)S^2D - (S^2 + n_3^2D^2)n_2^2D \nonumber \\
&= &S^2 - n_3^2D^2 + (1 + n_3^2)(D^2 + 2D)D - (S^2 + n_3^2D^2)n_2^2D \nonumber \\
&= &S^2 + n_3^2D^2 + D^3 + 2D^2 + n_3^2D^3 - (S^2 + n_3^2D^2)n_2^2D \nonumber \\
&= &S^2 + n_3^2D^2 + (S^2 + n_3^2D^2)D - (S^2 + n_3^2D^2)n_2^2D, \label{4.2.14} \ea
which is proportional to $C - n_2^2D$.
For the element $m_{32}$,
\ba s_{21} + \de s_2c_1 &= &{2n_3DS \over S^2 + n_3^2D^2}
+ (n_1^2 + n_2^2)D {(n_2S + n_1n_3D)(n_1S + n_2n_3D) \over [(n_1^2 + n_2^2)(S^2 + n_3^2D^2)]} \nonumber \\
&= &{2n_3DS \over S^2 + n_3^2D^2} + {[n_1n_2S^2 + (n_1^2 + n_2^2)n_3DS + n_1n_2n_3^2D^2]D \over S^2 + n_3^2D^2}. \label{4.2.15} \ea
The numerator in Eq. (\ref{4.2.15}) is
\ba &&[2D + (1 + n_3^2)D^2]n_3S + (S^2 + n_3^2D^2)n_1n_2D \nonumber \\
&= &[D(D + 2) + n_3^2D^2]n_3S + (S^2 + n_3^2D^2)n_1n_2D \nonumber \\
&= &(S^2 + n_3^2D^2)n_3S + (S^2 + n_3^2D^2)n_1n_2D, \label{4.2.16} \ea
which is proportional to $n_3S + n_1n_2D$. [In Eqs. (\ref{4.2.14}) and (\ref{4.2.16}), the identities mentioned before Eq. (\ref{4.2.7}) were used repeatedly.] The proofs of the equivalences of the formulas for $m_{23}$ and $m_{33}$ are similar.
Not only does the preceding analysis show that Eqs. (\ref{2.3.12}) and (\ref{4.2.4}) are equivalent, but it is also a constructive proof of the former equation. (An elegant, but abstract, proof is provided in App. E.)

Conversely, suppose that $\ga$, $\th_1$ and $\th_2$ are specified. Then, by comparing the traces of matrices (\ref{2.3.12}) and (\ref{4.2.4}), one finds that
\be C = [\ga - 1 + (\ga + 1)c_{21}]/2. \label{4.2.17} \ee
With $C$ known, so also are $D = C - 1$, $S = (C^2 - 1)^{1/2}$ and $k = \log(C + S)$. By adding and subtracting pairs of the off-diagonal entries, one finds that
\be n_1 = u(c_1 + c_2)/2S, \ \ n_2 \ = \ u(s_1 + s_2)/2S, \ \ n_3 \ = \ (\ga + 1)s_{21}/2S. \label{4.2.18} \ee
In Eqs. (\ref{4.2.4}) and (\ref{4.2.18}), changing the sign of $S$ is equivalent to changing the signs of $n_i$, so one can assume that $S > 0$ without loss of generality.
The inverse formulas for this $3 \times 3$ matrix are more complicated than the formulas for the $2 \times 2$ matrices. Nonetheless, they exist.

The proof that $M^{-1} = e^{-G}$ for transformations is similar to the proof for rotations ($G^3 = k^2G$ and $D = C - 1$). Changing the sign of $G$ is equivalent to changing the sign of $S$. Hence, the inverse matrix
\ba M^{-1} &= &\left[\begin{array}{ccc} C + n_3^2D & -n_1S + n_2n_3D & -n_2S - n_1n_3D \\
-n_1S - n_3n_2D & C - n_2^2D & n_3S + n_1n_2D \\
-n_2S + n_3n_1D & -n_3S + n_2n_1D & C - n_1^2D \end{array}\right]. \label{4.2.19} \ea
Notice that $M^{-1} = SM^tS$ (as it should do).

The derivation of formula (\ref{4.2.4}) was based on the assumption that $k_1^2 + k_2^2 - k_3^2 > 0$. In the opposite case, $k \rightarrow ik = i(k_3^2 - k_1^2 - k_2^2)^{1/2}$, $\sinh(k)/k \rightarrow \sin(k)/k$ and $[\cosh(k) - 1]/k^2 \rightarrow [1 - \cos(k)]/k^2$. With these changes, formulas (\ref{4.2.4}) and (\ref{4.2.19}) remain valid. ($G^3 \rightarrow -k^2G$, because the definition of $k$ changes.)

In this article, we only consider matrices with unit determinant, whose generators have zero trace. However, it is worth mentioning that the CH theorem also works for generators with nonzero trace. Let $G$ be an $n \times n$ matrix with $\tr(G) = nt$, and write $G = tI + H$. Then $\tr(tI) = \tr(G)$ and $\tr(H) = 0$, like the generators mentioned above. Furthermore, $tI$ commutes with $H$, so $\exp(tI + H) = \exp(t)\exp(H)$.

\newpage

\sec{5. Factorization of matrices}

In Secs. 3 and 4, specific formulas were derived for the matrices of the form
\be M = \exp(G_1k_1 + G_2k_2 + G_3k_3), \label{5.0.1} \ee
in which $G_1$, $G_2$ and $G_3$ act together. There are applications in which it is helpful to use the partial factorizations
\ba M &= &\exp(G_3l_3)\exp(G_1l_1 + G_2l_2) \nonumber \\
&= &\exp(G_1l_1' + G_2l_2')\exp(G_3l_3'), \label{5.0.2} \ea
in which $G_1$ and $G_2$ act together, and $G_3$ acts separately.
There are other applications in which it is helpful to use the full factorization.
\be M = \exp(G_3l_3)\exp(G_2l_2)\exp(G_1l_1), \label{5.0.3} \ee
in which $G_1$, $G_2$ and $G_3$ all act separately. In quantum optics, formulas such as (\ref{5.0.3}) are called disentanglement formulas \cite{wod85,ger01}.

In principle, one can determine factorizations (\ref{5.0.2}) and (\ref{5.0.3}) by Taylor expanding the exponential in Eq. (\ref{5.0.1}) and using the commutation relations [(\ref{2.1.7}), (\ref{2.2.5}), (\ref{2.2.15}), (\ref{2.3.5}) and (\ref{2.3.15})] to reorder the terms with $G_3$ to the left, and $G_1$ and $G_2$ on the right, or $G_3$ on the left, $G_2$ in the middle and $G_1$ on the right. This procedure is challenging and tedious. Nonetheless, it is clear from the outset that the factorization relations (formulas for $l_j$ in terms of $k_i$) depend on only the commutation relations: Groups with the same commutation relations have the same factorization formulas.

There are six partial factorizations, three with the single generator on the left and three with it on the right, and six full factorizations. It would be time-consuming to derive twelve factorizations (for each of five sets of generators). Fortunately, some guidance is provided by the Schmidt decomposition theorem \cite{hor13}.

Every complex matrix $M$ has the Schmidt decomposition $VDU^\d$, where $D$ is a diagonal matrix, and $U$ and $V$ are unitary matrices. The columns of $U$ (input Schmidt vectors) are the eigenvectors of $M^\d M$, the columns of $V$ (output Schmidt vectors) are the eigenvectors of $MM^\d$ and the diagonal entries of $D$ (Schmidt coefficients) are the square roots of the (common) eigenvalues of $M^\d M$ and $MM^\d$, which are non-negative.
If $U_i$ is an input vector, then $V_i =MU_i$ is the associated output vector, and conversely, if $V_i$ is an output vector, then $U_i = M^\d V_i$ is the associated input vector.
In terms of coefficients and vectors, $M = \tsum_i V_i\si_iU_i^\d$. One can multiply any pair of input and output vectors by an arbitrary phase factor without changing the decomposition.
If $M$ is hermitian, then the input and output vectors satisfy the same eigenvalue equation and the coefficients are the moduli of the eigenvalues of $M$, which are real. If the eigenvalue $\la_i \ge 0$, then $V_i = U_i$, whereas if $\la_i < 0$, then $V_i = -U_i$.

Likewise, every real matrix $M$ has the Schmidt decomposition $QDP^t$, where $D$ is diagonal and nonnegative, and $P$ and $Q$ are orthogonal \cite{hor13}. The columns of $P$ are the eigenvectors of $M^tM$, the columns of $Q$ are the eigenvectors of $MM^t$ and the entries of $D$ are the square roots of the eigenvalues of $M^tM$ and $MM^t$. If $M$ is symmetric, then $P = Q$ and the entries of $D$ are the moduli of the eigenvalues of $M$. One can change the signs of any pair of input and output vectors without changing the decomposition.

\subsec{5.1 Symplectic matrices}

The general form of a symplectic matrix was stated in Eq. (\ref{2.1.3}) and the generator form was stated in Eq. (\ref{3.1.4}). According to the real decomposition theorem, $M = QDP^t$, where $P$ and $Q$ are orthogonal (rotation) matrices. Hence, $M = (QP^t)(PDP^t)$. By using the alternative notation $P = R(\th_1) = R_1$ and $Q = R(\th_2) = R_2$, one can rewrite this equation in the form $M = R_{21}N_1$, where $N_1 = R_1DR_1^t$ represents a dilation with respect to axes inclined at $\th_1$ radians to the coordinate axes and $R_{21} = R_2R_1^t$ represents a rotation through $\th_{21} = \th_2 - \th_1$ radians \cite{mck25a}. Dilations and rotations are the building blocks of two-dimensional transformations, so this decomposition provides physical insight.

In \cite{mck25a}, we showed that the Schmidt product
\ba M &= &\left[\begin{array}{cc} c_2 & -s_2 \\ s_2 & c_2 \end{array}\right]
\left[\begin{array}{cc} C + S & 0 \\ 0 & C - S \end{array}\right]
\left[\begin{array}{cc} c_1 & s_1 \\ -s_1 & c_1\end{array}\right] \nonumber \\
&= &\left[\begin{array}{cc} c_2 & -s_2 \\ s_2 & c_2 \end{array}\right]
\left[\begin{array}{cc} (C+S)c_1 & (C+S)s_1 \\ (S - C)s_1 & (C-S)c_1 \end{array}\right] \nonumber \\
&= &\left[\begin{array}{cc} Cc_- + Sc_+ & Ss_+ - Cs_- \\ Ss_+ + Cs_- & Cc_- - Sc_+ \end{array}\right], \label{5.1.1} \ea
where  $C = \cosh(\la)$, $S = \sinh(\la)$, $c_i = \cos(\th_i)$, $s_i = \sin(\th_i)$ and $\th_\pm = \th_2 \pm \th_1$. In these formulas, $\la$ is the dilation parameter, and $\th_1$ and $\th_2$ are the input and output angles, respectively.
By comparing matrices (\ref{3.1.4}) and (\ref{5.1.1}), we found that
\be C^2 = C_0^2 + S_3^2, \ \ S^2 = S_1^2 + S_2^2, \label{5.1.2} \ee
where $C_0 = \cosh(k)$, $S_0 = \sinh(k)$, $S_i = S_0k_i/k$ and $k = (k_1^2 + k_2^2 - k_3^2)^{1/2}$. (Subscripts were added to the hyperbolic trigonometric functions in the generator form of the matrix to distinguish them from similar functions in the Schmidt form.)
We also found that the sum and difference angles are defined implicitly by the equations
\be \tan(\th_+) = S_2/S_1, \ \ \tan(\th_-) = S_3/C_0, \label{5.1.3} \ee
from which it follows that the input and output angles are defined by the equations
\be \tan(2\th_1) = {C_0S_2 - S_1S_3 \over C_0S_1 + S_2S_3}, \ \ 
\tan(2\th_2) = {C_0S_2 + S_1S_3 \over C_0S_1 - S_2S_3}. \label{5.1.4} \ee
Notice that $\th_+ = k_2/k_1$ is constant and $\th_-$ is nonzero if and only if $k_3$ is nonzero.
Equations (\ref{5.1.2}) and (\ref{5.1.4}) specify the decomposition parameters ($\la$, $\th_1$ and $\th_2$) in terms of the generator coefficients ($k_1$, $k_2$ and $k_3$).

With the Schmidt decomposition $M = R_2DR_1^t$ known, so also is alternative (dilation-rotation) decomposition $M = R_{21}N_1$. In the notation of Eq. (\ref{5.0.2}), the rotation matrix $R_{21} = \exp(G_3l_3)$, so $l_3 = \th_{21}$. Likewise, the symmetric matrix $N_1 = \exp(G_1l_1 + G_2l_2)$. It follows from Eq. (\ref{3.1.4}), with $n_3 = 0$, and Eq. (\ref{5.1.1}), with $\th_+ = 2\th_1$ and $\th_- = 0$, that $l_1 = l\cos(2\th_1) = lc_{11}$ and $l_2 = l\sin(2\th_1) = ls_{11}$, where $l = (l_1^2 + l_2^2)^{1/2} = \log(C + S)$. The preceding formulas specify $l_1$ -- $l_3$ in terms of $k_1$ -- $k_3$. One can verify these statements by checking that
\be L = \left[\begin{array}{cc} c_{21} & -s_{21} \\ s_{21} & c_{21} \end{array}\right]
\left[\begin{array}{cc} C + Sc_{11} &Ss_{11} \\ Ss_{11} & C - Sc_{11} \end{array}\right]. \label{5.1.5} \ee

\subsec{5.2 Unitary and indefinite unitary matrices}

Let $V$ be a unitary matrix whose general form was stated in Eq. (\ref{2.2.2}). Then, if $V$ acts on the unit vectors $[1, 0]^t$ and $[0, 1]^t$, it produces its own column vectors $V_1$ and $V_2$, respectively. The physical significance of $V$ is clear, so no further analysis is needed. This statement is bourne out by the Schmidt decomposition $V = VII^t$.

Now let $M$ be an indefinite unitary matrix, whose general form was stated in Eq. (\ref{2.2.12}) and whose generator form was stated in Eq. (\ref{3.3.4}). Then, for the special case in which $\mu$ and $\nu$ are real, it is easy to verify that
\be M = {1 \over 2^{1/2}} \left[\begin{array}{cc} 1 & -1 \\ 1 & 1 \end{array}\right]
\left[\begin{array}{cc} \mu + \nu & 0 \\ 0 & \mu - \nu \end{array}\right]
{1 \over 2^{1/2}} \left[\begin{array}{cc} 1 & 1 \\ -1 & 1 \end{array}\right]. \label{5.2.1} \ee
The third matrix in Eq. (\ref{5.2.1}) resolves the vector on which it acts into sum and difference vectors, the components of which are called the sum and difference amplitudes. The second matrix increases (stretches) the sum amplitude and decreases (squeezes) the difference amplitude, and the first matrix projects the dilated amplitudes onto the sum and difference vectors.
For the general case in which $\mu$ and $\nu$ are complex, let $\ph_s = (\ph_\mu + \ph_\nu)/2$ and $\ph_d = (\ph_\nu - \ph_\mu)/2$ be the sum and difference phases. Then it is easy to verify that
\be M = {1 \over 2^{1/2}} \left[\begin{array}{cc} e^{i\ph_s} & -e^{i\ph_s} \\ e^{-i\ph_s} & e^{-i\ph_s} \end{array}\right]
\left[\begin{array}{cc} |\mu| + |\nu| & 0 \\ 0 & |\mu| - |\nu| \end{array}\right]
{1 \over 2^{1/2}} \left[\begin{array}{cc} e^{-i\ph_d} & e^{i\ph_d} \\ -e^{-i\ph_d} & e^{-i\ph_d} \end{array}\right]. \label{5.2.2} \ee
Decomposition (\ref{5.2.2}) is similar to decomposition (\ref{5.2.1}). The main differences are that the input and output vectors are complex and distinct \cite{mck13}.

By comparing Eqs. (\ref{2.2.12}) and (\ref{3.3.4}), one finds that the dilation parameters
\be |\mu|^2 = C_0^2 + S_3^2,\ \ |\nu|^2 = S_1^2 + S_2^2, \label{5.2.3} \ee
where $C_0 = \cosh(k)$, $S_0 = \sinh(k)$, $S_i = S_0k_i/k$ and $k = (k_1^2 + k_2^2 - k_3^2)^{1/2}$. (Subscripts were added to the hyperbolic trigonometric functions in the generator form of the matrix.)
One also finds that the phases
\be \tan(\ph_\mu) = S_3/C_0, \ \ \tan(\ph_\nu) = S_2/S_1, \label{5.2.4} \ee
from which it follows that the sum and difference phases
\be \tan(\ph_\mu + \ph_\nu) = {C_0S_2 + S_1S_3 \over C_0S_1 - S_2S_3}, \ \ 
\tan(\ph_\nu - \ph_\mu) = {C_0S_2 - S_1S_3 \over C_0S_1 + S_2S_3}. \label{5.2.5} \ee
Equations (\ref{5.2.3}) and (\ref{5.2.5}) specify the decomposition parameters ($|\mu|$, $\ph_s$ and $\ph_d$) in terms of the generator coefficients ($k_1$, $k_2$ and $k_3$). The reason for the similarities between Eqs. (\ref{5.1.2}) and (\ref{5.1.4}), and Eqs. (\ref{5.2.3}) and (\ref{5.2.5}), will be explained in Sec. 6.

In the context of matrices, the Schmidt decomposition provides the required physical insight. No other factorizations are necessary. However, in the context of quantum operators, factorization (\ref{5.0.3}) facilitates studies of quantum evolution in the Schr\"odinger picture. This important factorization will be discussed in a future tutorial.

\subsec{5.3 Orthogonal and indefinite orthogonal matrices}

Let $R$ be a rotation matrix, whose general form was stated in Eq. (\ref{2.3.2}). Then the columns of $R$ are the images of the unit vectors $[1, 0, 0]^t$, $[0,1, 0]^t$ and $[0, 0, 1]^t$. The physical significance of $R$ is clear and the Schmidt decomposition, $R = RII^t$, is trivial.
Exponentiation produces the canonical matrix (\ref{4.1.7}), so there is no critical need for factorization. However, one might ask if one can reproduce an arbitrary rotation by a sequence of rotations about the $x$, $y$ and $z$ axes, especially if one is interested in computer animation \cite{fbe10} or spacecraft dynamics \cite{rui13}.

By exponentiating generators (\ref{4.1.1}) separately, one finds that
\be R_1 = \left[\begin{array}{ccc} 1 & 0 & 0 \\ 0 & c_1 & -s_1 \\ 0 & s_1 & c_1 \end{array}\right], \ \ 
R_2 = \left[\begin{array}{ccc} c_2 & 0 & s_2 \\ 0 & 1 & 0 \\ -s_2 & 0 & c_2 \end{array}\right], \ \ 
R_3 = \left[\begin{array}{ccc} c_3 & -s_3 & 0 \\ s_3 & c_3 & 0 \\ 0 & 0 & 1 \end{array}\right], \label{5.3.1} \ee
where $c_i = \cos\th_i$ and $s_i = \sin\th_i$. Hence,
\ba R_3R_2R_1 &= &\left[\begin{array}{ccc} c_3 & -s_3 & 0 \\ s_3 & c_3 & 0 \\ 0 & 0 & 1 \end{array}\right]
\left[\begin{array}{ccc} c_2 & 0 & s_2 \\ 0 & 1 & 0 \\ -s_2 & 0 & c_2 \end{array}\right]
\left[\begin{array}{ccc} 1 & 0 & 0 \\ 0 & c_1 & -s_1 \\ 0 & s_1 & c_1 \end{array}\right] \nonumber \\
&= &\left[\begin{array}{ccc} c_3 & -s_3 & 0 \\ s_3 & c_3 & 0 \\ 0 & 0 & 1 \end{array}\right]
\left[\begin{array}{ccc} c_2 & s_2s_1 & s_2c_1 \\ 0 & c_1 & -s_1 \\ -s_2 & c_2s_1 & c_2c_1 \end{array}\right] \nonumber \\
&= &\left[\begin{array}{ccc} c_3c_2 & c_3s_2s_1 - s_3c_1 & c_3s_2c_1 + s_3s_1 \\
s_3c_2 & s_3s_2s_1 + c_3c_1 & s_3s_2c_1 - c_3s_1 \\ -s_2 & c_2s_1 & c_2c_1 \end{array}\right]. \label{5.3.2} \ea
By comparing the first columns and last rows of matrices (\ref{4.1.7}) and (\ref{5.3.2}), one finds that
\be t_1 = {n_1s_0 + n_2n_3d_0 \over c_0 + n_3^2d_0}, \ \ 
s_2 = n_2s_0 - n_1n_3d_0, \ \ 
t_3 = {n_3s_0 + n_1n_2d_0 \over c_0 + n_1^2d_0}, \label{5.3.3b} \nonumber \ee
where $t_i = \tan\th_i$, $c_0 = \cos(k)$, $s_0 = \sin(k)$, $d_0 = 1 - c_0$, $n_i = k_i/k$ and $k = (k_1^2 + k_2^2 + k_3^2)^{1/2}$. (Subscripts were added to the trigonometric functions in the canonical form of the matrix.)
Equations (\ref{5.3.3b}) specify the product angles ($\th_1$, $\th_2$ and $\th_3$) in terms of the generator coefficients ($k_1$, $k_2$ and $k_3$) or, equivalently, the axis-angle parameters ($n_i$ and $\th = k$).

We repeated the calculation for the ordering $R_2R_1R_3$, which is a cyclic permutation of the ordering $R_3R_2R_1$, and obtained the product-angle equations
\be t_3 = {n_3s_0 + n_1n_2d_0 \over c_0 + n_2^2d_0}, \ \ 
s_1 = n_1s_0 - n_2n_3d_0, \ \ 
t_2 = {n_2s_0 + n_1n_3d_0 \over c_0 + n_3^2d_0}. \label{5.3.3d} \nonumber \ee
The similarities of Eqs. (\ref{5.3.3b}) and (\ref{5.3.3d}) are clear.
These results are examples of operator ordering [Eq. (\ref{5.0.3})]. In the generator form of the rotation matrix [Eq. (\ref{4.1.7})], $G_1$, $G_2$ and $G_3$ appear together, whereas in the product forms, $G_1$, $G_2$ and $G_3$ appear separately, in chosen orders.

Now let $L$ be a Lorentz matrix, whose general form was stated in Eq. (\ref{2.3.12}) and whose generator form was stated in Eq. (\ref{4.2.4}).
The Schmidt-like product
\ba L &= &\left[\begin{array}{ccc} 1 & 0 & 0 \\ 0 & c_2 & -s_2 \\ 0 & s_2 & c_2 \end{array}\right]
\left[\begin{array}{ccc} C & S & 0 \\ S & C & 0 \\ 0 & 0 & 1 \end{array}\right]
\left[\begin{array}{ccc} 1 & 0 & 0 \\ 0 & c_1 & s_1 \\ 0 & -s_1 & c_1 \end{array}\right] \nonumber \\
&= &\left[\begin{array}{ccc} 1 & 0 & 0 \\ 0 & c_2 & -s_2 \\ 0 & s_2 & c_2 \end{array}\right]
\left[\begin{array}{ccc} C & Sc_1 & Ss_1 \\ S & Cc_1 & Cs_1 \\ 0 & -s_1 & c_1 \end{array}\right] \nonumber \\
&= &\left[\begin{array}{ccc} C & Sc_1 & Ss_1 \\ Sc_2 & c_{21} + Dc_2c_1 & -s_{21} + Dc_2s_1 \\ Ss_2 & s_{21} + Ds_2c_1 & c_{21} + Ds_2s_1 \end{array}\right], \label{5.4.1} \ea
where $C = \cosh(\la)$, $c_i= \cos(\th_i)$ and $\th_{21} = \th_2 - \th_1$. The definitions of $S$ and $s_i$ are similar. Formula (\ref{5.4.1}) is like formula (\ref{2.3.12}), with $\ga = C$, $\de = D$ and $u = S$. 
By comparing Eqs. (\ref{4.2.4}) and (\ref{5.4.1}), one finds that
\be C = C_0 + n_3^2D_0, \ \ S = [(n_1^2 + n_2^2)(S_0^2 + n_3^2D_0^2)]^{1/2}, \label{5.4.2} \ee
%
from which it follows that $D = (n_1^2 + n_2^2)D_0$. (Once again, subscripts were added to the hyperbolic trigonometric functions in the generator form of the matrix.) One also finds that
\be \tan(\th_1) = {n_2S_0 - n_1n_3D_0 \over n_1S_0 + n_2n_3D_0}, \ \ 
\tan(\th_2) = {n_2S_0 + n_1n_3D_0 \over n_1S_0 - n_2n_3D_0}. \label{5.4.3} \ee
By combining Eqs. (\ref{5.4.3}) with trigonmetric identities, or by adding and subtracting elements of the lower-right block of matrices (\ref{4.2.4}) and (\ref{5.4.1}), one finds that
\be \tan(\th_+) = {2n_1n_2 \over n_1^2 - n_2^2}, \ \ 
\tan(\th_-) = {2n_3D_0S_0 \over S_0^2 - n_3^2D_0^2}, \label{5.4.4} \ee
where $\th_\pm = \th_2 \pm \th_1$. Equations (\ref{5.4.2}) and (\ref{5.4.3}) specify the decomposition parameters ($\ga$, $\th_1$ and $\th_2$) in terms of the generator coefficients ($k_1$, $k_2$ and $k_3$).

Decomposition (\ref{5.4.1}) differs from the standard Schmidt decomposition in two important ways: First, the first and third matrices on the right side represent two-dimensional rotations (about the $t$ axis, in the $xy$ plane), which are special cases of orthogonal transformations (three-dimensional rotations). Second, the second matrix represents a boost, not a dilation. It is easy to verify that a boost of the variables $t$ and $x$ is equivalent to a dilation of the sum and difference variables $t \pm x$ [Eqs. (\ref{2.2.12}) and (\ref{5.2.1})]. However, if one were to use the latter variables, then the decomposition would involve three-dimensional rotations, which are harder to visualize than two-dimensional ones.

With the Schmidt-like decomposition $L = R_2BR_1^t$ known, so also is alternative (boost-rotation) decomposition $L = R_{21}N_1$, where $N_1 = R_1BR_1^t$. In the notation of Eq. (\ref{5.0.2}), the rotation matrix $R_{21} = \exp(G_3l_3)$, so $l_3 = \th_{21}$. Likewise, the symmetric matrix $N_1 = \exp(G_1l_1 + G_2l_2)$. It follows from Eqs. (\ref{4.2.5}) and (\ref{5.4.1}) that $l_1 = lc_1$ and $l_2 = ls_1$, where $l = (l_1^2 + l_2^2)^{1/2} = \log(C + S)$. The preceding formulas specify $l_1$ -- $l_3$ in terms of $k_1$ -- $k_3$. One can verify these statements by checking that
\be L = \left[\begin{array}{ccc} 1 & 0 & 0 \\ 0 & c_{21} & -s_{21} \\ 0 & s_{21} & c_{21} \end{array}\right]
\left[\begin{array}{ccc} C & Sc_1 & Ss_1 \\ Sc_1 & 1 + Dc_1^2 & Dc_1s_1 \\ Cs_1 & Ds_1c_1 & 1 + Ds_1^2 \end{array}\right]. \label{5.3.11} \ee

\newpage

\sec{6. Isomorphisms and commutation relations}

In this section, we explain why the commutation relations matter and demonstrate the importance of isomorphisms.
Let $A$, $B$ and $C$ be members of matrix group one, $X$, $Y$ and $Z$ be members of group two, and suppose that there is a one-to-one relation between the groups. Then every object matrix $A$ has one image matrix $X$, and every image matrix $Y$ has one object $B$, of which it is the image. This relation between the groups is called an isomorphism if it preserves multiplication: $AB = C$ in group one if and only if $XY = Z$ in group two, where $Z$ is the image of $C$ \cite{taw95}. The matrices in group one need not have the same dimensions as those in group two. (In fact, the members of group two need not be matrices and their binary operation need not be matrix multiplication. The relation between the groups is isomorphic if it preserves the rules of binary operation.) If two groups are isomorphic, they have the same structure. For example, if $B = A^{-1}$, then $Y = X^{-1}$. If $A$, $B$ and $C$ form a subgroup of group one, then $X$, $Y$ and $Z$ form a subgroup of group two. One can establish results for the simpler-to-analyze group (for example, $2 \times 2$ matrices) and know, without further effort, that they are also true for the harder-to-analyze group (for example, $3 \times 3$ matrices).

Let $G_i$ be a generator of group one and $H_i$ be a generator of group two. (We assume that both groups have three generators.) Then every matrix $A$ can be written as the exponential $\exp(\al_1G_1 + \al_2G_2 + \al_3G_3)$. It can also be written as $\exp(\al_3'G_3)\exp(\al_2'G_2)\exp(\al_1'G_1)$. In this discussion, we will use the second form and omit the primes. Consider the product
\ba BA &= &\exp(\bt_3G_3)\exp(\bt_2G_2)\exp(\bt_1G_1) \nonumber \\
&&\times \exp(\al_3G_3)\exp(\al_2G_2)\exp(\al_1G_1). \label{6.0.1} \ea
By using the commutation relations, one can rewrite this matrix in the canonical form
\be C = \exp(\ga_3G_3)\exp(\ga_2G_2)\exp(\ga_1G_1), \label{6.0.2} \ee
where $\ga_k = g_k(\al_i, \bt_j)$. Likewise, one can write the product
\ba YX &= &\exp(\et_3H_3)\exp(\et_2H_2)\exp(\et_1H_1) \nonumber \\
&&\times \exp(\xi_3H_3)\exp(\xi_2H_2)\exp(\xi_1H_1) \label{6.0.3} \ea
in the canonical form
\be Z = \exp(\ze_3H_3)\exp(\ze_2H_2)\exp(\ze_1H_1), \label{6.0.4} \ee
where $\ze_k = h_k(\xi_i, \et_j)$. The decomposition formulas (for $\ga_k$ and $\ze_k$) depend solely on the commutation relations, so if the two sets of generators satisfy the same relations, the functions must be the same ($g_k = h_k$). Thus, the natural way to relate two groups whose generators satisfy the same commutation relations is to equate their generator coefficients ($\al_i = \xi_i$). This relation preserves the rules of multiplication ($AB = C$ if and only if $YX = Z$), so the two groups are isomorphic. This result shows the importance of commutation relations: Given two sets of generators, one only has to compare their commutation relations (which is easy to do) to determine whether (or not) the groups are isomorphic.

In the following subsections, we will discuss the relation between the matrices in Sp(2) and SU(1,1), which are real and complex $2 \times 2$ matrices, respectively, Sp(2) and SO(1,2), which are real $2 \times 2$ and $3 \times 3$ matrices, respectively, and SU(2) and SO(3), which are complex $2 \times 2$ and real $3 \times 3$ matrices, respectively. In these discussions, we will restate the main results of Secs. 3 -- 5, for convenience. (It is easier to compare equations when they are nearby.)

\subsec{6.1 Sp(2) and SU(1,1)}

Every member of Sp(2) can be written in the form
\be M = \left[\begin{array}{cc} C_0 + S_1 & S_2 - S_3 \\ S_2 + S_3 & C_0 - S_1 \end{array}\right], \label{6.1.1} \ee
where $C_0 = \cosh(k)$, $S_0 = \sinh(k)$, $k = (k_1^2 + k_2^2 - k_3^2)^{1/2}$ and $S_i = S_0k_i/k$. Likewise, every member of  SU(1,1) can be written in the form
\be M = \left[\begin{array}{cc} C_0 + iS_3 & S_1 + iS_2 \\ S_1 - iS_2 & C_0 - iS_3 \end{array}\right]. \label{6.1.2} \ee
The matrices in Eqs. (\ref{6.1.1}) and (\ref{6.1.2}) have the same generator coefficients, so there is a one-to-one relation between the matrices in the two groups. It only remains to show that the rules of multiplication are preserved.

Let $M$ and $M'$ be members of Sp(2). Then their product
\be M'' = \left[\begin{array}{cc} C_0' + S_1' & S_2' - S_3' \\ S_2' + S_3' & C_0' - S_1' \end{array}\right]
\left[\begin{array}{cc} C_0 + S_1 & S_2 - S_3 \\ S_2 + S_3 & C_0 - S_1 \end{array}\right]. \label{6.1.3} \ee
It is easy to verify that
\ba m_{11}'' &= &(C_0'C_0 + S_1'S_1 + S_2'S_2 - S_3'S_3) + (C_0'S_1 + S_1'C_0 + S_2'S_3 - S_3'S_2), \label{6.1.4} \\
m_{12}'' &= &(C_0'S_2 + S_2'C_0 + S_3'S_1 - S_1'S_3) - (C_0'S_3 + S_3'C_0 + S_2'S_1 - S_1'S_2). \label{6.1.5} \ea
According to Eq. (\ref{6.1.1}), the bracketed terms in Eqs. (\ref{6.1.4}) and (\ref{6.1.5}) represent $C_0''$, $S_1''$, $S_2''$ and $S_3''$, respectively. Now let $M$ and $M'$ be members of SU(1,1). Then their product
\be M'' = \left[\begin{array}{cc} C_0' + iS_3' & S_1' + iS_2' \\ S_1' - iS_2' & C_0' - iS_3' \end{array}\right]
\left[\begin{array}{cc} C_0 + iS_3 & S_1 + iS_2 \\ S_1 - iS_2 & C_0 - iS_3 \end{array}\right]. \label{6.1.6} \ee
It is easy to verify that
\ba m_{11}'' &= &(C_0'C_0 + S_1'S_1 + S_2'S_2 - S_3'S_3) + i(C_0'S_3 + S_3'C_0 + S_2'S_1 - S_1'S_2), \label{6.1.7} \\
m_{12}'' &= &(C_0'S_1 + S_1'C_0 + S_2'S_3 - S_3'S_2) + i(C_0'S_2 + S_2'C_0 + S_3'S_1 - S_1'S_3). \label{6.1.8} \ea
According to Eq. (\ref{6.1.2}), the bracketed terms in Eqs. (\ref{6.1.7}) and (\ref{6.1.8}) represent $C_0''$, $S_3''$, $S_1''$ and $S_2''$, respectively. The formulas for the components of the product matrices are identical, so the rules of multiplication are preserved. Hence, Sp(2) and SU(1,1) are isomorphic.

Isomorphism manifests itself in matrix decomposition. If $M$ is in Sp(2), then it has the Schmidt decomposition $QDP^t$, where $D$ is diagonal, and $P$ and $Q$ are orthogonal (Sec. 5.1). Likewise, if $M$ is in SU(1,1), then it has the decomposition $VDU^\d$, where $U$ and $V$ are unitary (Sec. 5.2).
Let $M_i = \exp(G_ik_i)$. Then, for Sp(2), the fundamental matrices are
\be M_1 = \left[\begin{array}{cc} E_1 & 0 \\ 0 & E_1^{-1} \end{array}\right], \ \ \\
M_2 = \left[\begin{array}{cc} C_2 & S_2 \\ S_2 & C_2 \end{array}\right], \ \ \\
M_3 = \left[\begin{array}{cc} c_3 & -s_3 \\ s_3 & c_3 \end{array}\right], \ \ \label{6.1.11} \ee
where $E_1 = \exp(k_1)$, $C_2 = \cosh(k_2)$, $S_2 = \sinh(k_2)$, $c_3 = \cos(k_3)$ and $s_3 = \sin(k_3)$. $M_1$ is diagonal and $M_3$ is orthogonal.
In Sec. 5.1, it was shown that the Schmidt product
\ba M &= &\left[\begin{array}{cc} c_2 & -s_2 \\ s_2 & c_2 \end{array}\right]
\left[\begin{array}{cc} C + S & 0 \\ 0 & C - S \end{array}\right]
\left[\begin{array}{cc} c_1 & s_1 \\ -s_1 & c_1\end{array}\right] \nonumber \\
&= &\left[\begin{array}{cc} Cc_- + Sc_+ & Ss_+ - Cs_- \\ Ss_+ + Cs_- & Cc_- - Sc_+ \end{array}\right], \label{6.1.12} \ea
where  $C = \cosh(\la)$, $c_i = \cos(\th_i)$ and $\th_\pm = \th_2 \pm \th_1$. The definitions of $S$ and $s_i$ are similar.
The dilation parameter is specified by the equations
\be C^2 = C_0^2 + S_3^2, \ \ S^2 = S_1^2 + S_2^2. \label{6.1.13} \ee
The sum and difference angles are specified implicitly by the equations
\be \tan(\th_+) = S_2/S_1, \ \ \tan(\th_-) = S_3/C_0, \label{6.1.14} \ee
and the input and output angles are specified by the equations
\be \tan(2\th_1) = {C_0S_2 - S_1S_3 \over C_0S_1 + S_2S_3}, \ \ 
\tan(2\th_2) = {C_0S_2 + S_1S_3 \over C_0S_1 - S_2S_3}. \label{6.1.15} \ee

For SU(1,1), the fundamental matrices are
\be M_1 = \left[\begin{array}{cc} C_1 & S_1 \\ S_1 & C_1 \end{array}\right], \ \ 
M_2 = \left[\begin{array}{cc} C_2 & iS_2 \\ -iS_2 & C_2 \end{array}\right], \ \ 
M_3 = \left[\begin{array}{cc} e_3 & 0 \\ 0 & e_3^* \end{array}\right], \ \ \label{6.1.21} \ee
where $C_i = \cosh(k_i)$, $S_i = \sinh(k_i)$ and $e_3 = \exp(ik_3)$. $M_1$ is not diagonal (it represents a boost rather than a dilation), but $M_3$ is unitary (it represents a differential phase shift).
The Schmidt-like product
\ba M &= &\left[\begin{array}{cc} e_2 & 0 \\ 0 & e_2^* \end{array}\right]
\left[\begin{array}{cc} C & S \\ S & C \end{array}\right]
\left[\begin{array}{cc} e_1^* & 0 \\ 0 & e_1 \end{array}\right] \nonumber \\
&= &\left[\begin{array}{cc} e_2 & 0 \\ 0 & e_2^* \end{array}\right]
\left[\begin{array}{cc} Ce_1^* & Se_1 \\ Se_1^* & Ce_1 \end{array}\right] \nonumber \\
&= &\left[\begin{array}{cc} Ce_2e_1^* & Se_2e_1 \\ Se_2^*e_1^* & Ce_2^*e_1 \end{array}\right], \label{6.1.22} \ea
where $e_i = \exp(i\th_i)$. By comparing Eqs. (\ref{6.1.2}) and (\ref{6.1.22}), one finds that
\be C^2 = C_0^2 + S_3^2, \ \ S^2 = S_1^2 + S_2^2. \label{6.1.23} \ee
One also finds that
\be \tan(\th_+) = S_2/S_1, \ \ \tan(\th_-) = S_3/C_0, \label{6.1.24} \ee
from which it follows that
\be \tan(2\th_1) = {C_0S_2 - S_1S_3 \over C_0S_1 + S_2S_3}, \ \ 
\tan(2\th_2) = {C_0S_2 + S_1S_3 \over C_0S_1 - S_2S_3}. \label{6.1.25} \ee
By comparing Eqs. (\ref{5.2.2}) and (\ref{6.1.22}), one finds that $\th_2 = (\ph_\mu + \ph_\nu)/2 = \ph_s$ and $\th_1 = (\ph_\nu - \ph_\mu)/2 = \ph_d$, so Eqs. (\ref{6.1.23}) and (\ref{6.1.25}) are equivalent to Eqs. (\ref{5.2.3}) and (\ref{5.2.5}), respectively. They are also identical to Eqs. (\ref{6.1.13}) and (\ref{6.1.15}), respectively. Provided that one uses the associated fundamental matrices in Sp(2) and SU(1,1), the decomposition formulas are identical. (In Sec. 5.2, it was shown that a boost can be decomposed into a dilation and two unitary transformations, so the Schmidt-like decomposition is a Schmidt decomposition written in a different way.)

In retrospect, we should have anticipated this result. If two groups are isomorphic, then $BA = C$ if and only if $YX = Z$. By extension, $CBA = D$ if and only if $YXW = Z$. Hence, if $M_s = M_{3s}(\th_2)M_{1s}(\la)M_{3s}^t(\th_1)$, where $s$ denotes symplectic, then $M_u = M_{3u}(\th_2)M_{1u}(\la)M_{3u}^\d(\th_1)$, where $u$ denotes indefinite unitary. This relation is true, even though $M_{1s}$ is a dilation, whereas $M_{1u}$ is a boost, and $M_{3s}$ is a rotation, whereas $M_{3u}$ is a phase shift.
A similar statement can be made about the decompositions of matrices in Sp(2) and SO(1,2).

\subsec{6.2 Sp(2) and SO(1,2)}

The general form of a matrix in Sp(2) was stated in Eq. (\ref{6.1.1}) and the Schmidt decomposition was specified by Eqs. (\ref{6.1.12}), (\ref{6.1.13}) and (\ref{6.1.15}).
Every member of SO(1,2) can be written in the form
\ba L &= &\left[\begin{array}{ccc} C_0 + n_3^2D_0 & n_1S_0 + n_2n_3D_0 & n_2S_0 - n_1n_3D_0 \\
n_1S_0 - n_3n_2D_0 & C_0 - n_2^2D_0 & -n_3S_0 + n_1n_2D_0 \\
n_2S_0 + n_3n_1D_0 & n_3S_0 + n_2n_1D_0 & C_0 - n_1^2D_0 \end{array}\right], \label{6.2.1} \ea
where $C_0 = \cosh(k)$, $D_0 = C_0 - 1$, $S_0 = \sinh(k)$, $k = (k_1^2 + k_2^2 - k_3^2)^{1/2}$ and $n_i = k_i/k$.
For SO(1,2), the fundamental matrices are
\be L_1 = \left[\begin{array}{ccc} C_1 & S_1 & 0 \\ S_1 & C_1 & 0 \\ 0 & 0 & 1 \end{array}\right], \ \ 
L_2 = \left[\begin{array}{ccc} C_2 & 0 & S_2 \\ 0 & 1 & 0 \\ S_2 & 0 & C_2 \end{array}\right], \ \ 
L_3 = \left[\begin{array}{ccc} 1 & 0 & 0 \\ 0 & c_3 & -s_3 \\ 0 & s_3 & c_3 \end{array}\right], \ \ \label{6.2.2} \ee
where $C_i = \cosh(k_i)$, $S_i = \sinh(k_i)$, $c_3 = \cos(k_3)$ and $s_3 = \sin(k_3)$. $L_1$ and $L_2$ are not diagonal (they represent boosts, not dilations), but $L_3$ is orthogonal (and represents a rotation). In Sec. 5.3, it was shown that the Schmidt-like product
\ba L &= &\left[\begin{array}{ccc} 1 & 0 & 0 \\ 0 & c_2 & -s_2 \\ 0 & s_2 & c_2 \end{array}\right]
\left[\begin{array}{ccc} C & S & 0 \\ S & C & 0 \\ 0 & 0 & 1 \end{array}\right]
\left[\begin{array}{ccc} 1 & 0 & 0 \\ 0 & c_1 & s_1 \\ 0 & -s_1 & c_1 \end{array}\right] \nonumber \\
&= &\left[\begin{array}{ccc} C & Sc_1 & Ss_1 \\ Sc_2 & c_{21} + Dc_2c_1 & -s_{21} + Dc_2s_1 \\ Ss_2 & s_{21} + Ds_2c_1 & c_{21} + Ds_2s_1 \end{array}\right], \label{6.2.3} \ea
where $C = \cosh(\la)$, $c_i= \cos(\th_i)$ and $\th_{21} = \th_2 - \th_1$. The definitions of $S$ and $s_i$ are similar.
The dilation parameter is specified by the equations
\be C = C_0 + n_3^2D_0, \ \ S = [(n_1^2 + n_2^2)(S_0^2 + n_3^2D_0^2)]^{1/2}, \label{6.2.4} \ee
%
and the input and output angles are specified implicitly by the equations
\be \tan(\th_1) = {n_2S_0 - n_1n_3D_0 \over n_1S_0 + n_2n_3D_0}, \ \ 
\tan(\th_2) = {n_2S_0 + n_1n_3D_0 \over n_1S_0 - n_2n_3D_0}. \label{6.2.5} \ee

At first glance, Eqs. (\ref{6.2.4}) and (\ref{6.2.5}) look nothing like their counterparts, Eqs. (\ref{6.1.13}) and (\ref{6.1.15}), respectively. However, matrix (\ref{6.2.1}) is the exponential of canonical generators, whose commutation relations have coefficients of $\pm 1$ on their right sides. In contrast, matrix (\ref{6.1.1}) is the exponential of generators, whose commutation relations have coefficients of $\pm 2$ on their right sides. In order to make a fair comparison between the results, one must replace the Sp(2) coefficients $k_i$ by $k_i/2$.

For hyperbolic trigonometric functions, the full- or half-angle formulas are $C_f= C_h^2 + S_h^2$ and $S_f = 2S_hC_h$,  or, equivalently, $C_h^2 = (C_f + 1)/2$ and $S_h^2 = (C_f - 1)/2 = D_f/2$, where $f$ and $h$ mean full and half, respectively. It follows from Eqs. (\ref{6.1.13}) that
\ba C^2 + S^2 &= &C_0^2 + (n_1^2 + n_2^2 + n_3^2)S_0^2 \nonumber \\
&= &(C_0^2 + S_0^2) + n_3^2(2S_0^2), \label{6.2.6} \\
(2SC)^2 &= &4(n_1^2 + n_2^2)S_0^2(C_0^2 + n_3^2S_0^2) \nonumber \\
&= &(n_1^2 + n_2^2)[(2S_0C_0)^2 + n_3^2(2S_0^2)^2]. \label{6.2.7} \ea
Equations (\ref{6.2.6}) and (\ref{6.2.7}) are equivalent to Eqs. (\ref{6.2.4}). It follows from Eqs. (\ref{6.1.15}) that
\be \tan(2\th_1) = {n_2(2S_0C_0) - n_1n_3(2S_0^2) \over n_1(2S_0C_0) + n_2n_3(2S_0^2)}, \ \ 
\tan(2\th_2) = {n_2(2S_0C_0) + n_1n_3(2S_0^2) \over n_1(2S_0C_0) - n_2n_3(2S_0^2)}. \label{6.2.8} \ee
Equations (\ref{6.2.8}) are equivalent to Eqs. (\ref{6.2.5}). Provided that one uses the associated fundamental matrices in Sp(2) and SO(1,2), the decomposition formulas are identical.

One also encounters Schmidt decompositions when one considers the product of two matrices. Let $M_1$ and $M_2$ be members of Sp(2). Then each matrix has the Schmidt decomposition $QDP^t$, where $D(\la)$ is a dilation, and $P(\ph)$ and $Q(\th)$ are rotations. Hence, the product matrix
\be M_2M_1 = Q_2D_2P_2^tQ_1D_1P_1^t = Q_2(D_2R_{21}^tD_1)P_1^t, \label{6.2.11} \ee
where $R_{21} = P_2Q_1^t$ is a (differential) rotation matrix. (Two-dimensional rotation matrices commute.) In Eq. (\ref{6.2.11}), $P_1$ and $Q_2$ represent rotations, the effects of which are simple. The key product is the intermediate matrix $M_3 = D_2R_{21}^tD_1$, which determines how the transformations interact. (Is the composite dilation stronger or weaker than the component dilations?)
Now let $L_1$ and $L_2$ be members of SO(1,2). Then each matrix has the Schmidt-like decomposition $QBP^t$, where $B(\ga)$ is a boost, and $P(\ph)$ and $Q(\th)$ are two-dimensional rotations. Hence, the product matrix
\be L_2L_1 = Q_2B_2P_2^tQ_1B_1P_1^t = Q_2(B_2R_{21}^tB_1)P_1^t, \label{6.2.12} \ee
where $R_{21} = P_2Q_1^t$ is a two-dimensional rotation matrix. Once again, the key product is the intermediate matrix $L_3 = B_2R_{21}^tB_1$.

For symplectic matrices, the intermediate matrix
\ba M_3 &= &\left[\begin{array}{cc} C_2 + S_2 & 0 \\ 0 & C_2 - S_2 \end{array}\right]
\left[\begin{array}{cc} c & s \\ -s & c \end{array}\right]
\left[\begin{array}{cc} C_1 + S_1 & 0 \\ 0 & C_1 - S_1 \end{array}\right] \nonumber \\
&= &\left[\begin{array}{cc} C_2 + S_2 & 0 \\ 0 & C_2 - S_2 \end{array}\right]
\left[\begin{array}{cc} c(C_1 + S_1) & s(C_1 - S_1) \\ -s(C_1 + S_1) & c(C_1 - S_1) \end{array}\right] \nonumber \\
&= &\left[\begin{array}{cc} (C_2 + S_2)(C_1 + S_1)c & (C_2 + S_2)(C_1 - S_1)s \\ -(C_2 - S_2)(C_1 + S_1)s & (C_2 - S_2)(C_1 - S_1)c \end{array}\right], \label{6.2.13} \ea
where $C_i = \cosh(\la_i)$ and $c = \cos(\ph_2 - \th_1)$. The definitions of $S_i$ and $s$ are similar. Notice that $\la_i$, $C_i$ and $S_i$ are alternative measures of the dilation strengths.
In \cite{mck25a}, we showed that matrix (\ref{6.2.13}) has the decomposition $M_3 = Q_3(\th_3)D_3(\la_3)P_3^t(\ph_3)$, where the dilation parameters
\ba C_3^2 &= &C_2^2C_1^2 + S_2^2S_1^2 + 2S_2C_2S_1C_1(c^2 - s^2), \label{6.2.14} \\
S_3^2 &= &S_2^2C_1^2 + C_2^2S_1^2 + 2S_2C_2S_1C_1(c^2 - s^2), \label{6.2.15} \ea
and the sum and difference angles are specified implicitly by the equations
\be \tan(\th_3 + \ph_3) = {(S_2C_1 - C_2S_1)s \over (S_2C_1 + C_2S_1)c}, \ \ 
\tan(\th_3 - \ph_3) =  -{(C_2C_1 - S_2S_1)s \over (C_2C_1 + S_2S_1)c}. \label{6.2.16} \ee
It follows from Eqs. (\ref{6.2.16}) that the input and output angles are specified by the equations
\ba \tan(2\ph_3) &= &{2S_2C_2sc \over (C_2^2 + S_2^2)S_1C_1 + S_2C_2(C_1^2 + S_1^2)(c^2 - s^2)}, \label{6.2.17} \\
\tan(2\th_3) &= &{-2S_1C_1sc \over S_2C_2(C_1^2 + S_1^2) + (C_2^2 + S_2^2)S_1C_1(c^2 - s^2)}. \label{6.2.18} \ea
Equations (\ref{6.2.14}), (\ref{6.2.15}), (\ref{6.2.17}) and (\ref{6.2.18}) are the multiplication rules for symplectic matrices, written in terms of the Schmidt parameters $\la$, $\ph$ and $\th$.

As we explained above, these parameters (arguments) are actually half arguments. In terms of full arguments, the dilation equation is
\be C_3 = C_2C_1 + S_2S_1c, \label{6.2.19} \ee
and the angle equations are
\be \tan(\ph_3) = {S_2s \over C_2S_1 + S_2C_1c}, \ \ 
\tan(\th_3) = {-S_1s \over S_2C_1 + C_2S_1c}. \label{6.2.20} \ee
Remarkably, the full-argument formulas are simpler than the half-argument formulas.

For Lorentz matrices, the intermediate matrix
\ba L_3 &= &\left[\begin{array}{ccc} \ga_2 & u_2 & 0 \\ u_2 & \ga_2 & 0 \\ 0 & 0 & 1 \end{array}\right]
\left[\begin{array}{ccc} 1 & 0 & 0 \\ 0 & c & s \\ 0 & -s & c \end{array}\right]
\left[\begin{array}{ccc} \ga_1 & u_1 & 0 \\ u_1 & \ga_1 & 0 \\ 0 & 0 & 1 \end{array}\right] \nonumber \\
&= &\left[\begin{array}{ccc} \ga_2 & u_2 & 0 \\ u_2 & \ga_2 & 0 \\ 0 & 0 & 1 \end{array}\right]
\left[\begin{array}{ccc} \ga_1 & u_1 & 0 \\ cu_1 & c\ga_1 & s \\ -su_1 & -s\ga_1 & c \end{array}\right] \nonumber \\
&= &\left[\begin{array}{ccc} \ga_2\ga_1 + u_2u_1c & \ga_2u_1 + u_2\ga_1c & u_2s \\ u_2\ga_1 + \ga_2u_1c & u_2u_1 + \ga_2\ga_1c & \ga_2s \\ -u_1s & -\ga_1s & c\end{array}\right], \label{6.2.21} \ea
where $\ga_i$ and $u_i = (\ga_i^2 - 1)^{1/2}$ are the energy and momentum parameters, respectively, $c = \cos(\ph_2 - \th_1)$ and $s = \sin(\ph_2 - \th_1)$. This matrix is the product of Lorentz matrices, so it is also a Lorentz matrix, with the decomposition $L_3 = Q_3(\th_3)B_3(\ga_3)P_3^t(\ph_3)$. According to Eq. (\ref{2.3.12}), the first row of the product matrix is $[\ga_3, u_3c_\ph, u_3s_\ph]$, where $c_\ph = \cos(\ph_3)$, and the first column is $[\ga_3, u_3c_\th, u_3s_\th]^t$, where $c_\th = \cos(\th_3)$. The definitions of $s_\ph$ and $s_\th$ are similar.
It follows from Eq. (\ref{6.2.21}) that the intermediate energy
\be \ga_3 = \ga_2\ga_1 + u_2u_1c. \label{6.2.22} \ee
It also follows that the intermediate angles are specified by the equations
\be \tan(\ph_3) = {u_2s \over \ga_2u_1 + u_2\ga_1c}, \ \ 
\tan(\th_3) = {-u_1s \over u_2\ga_1 + \ga_2u_1c}. \label{6.2.23} \ee
Equations (\ref{6.2.22}) and (\ref{6.2.23}) are the multiplication rules for Lorentz matrices, written in terms of the Schmidt-like parameters $\ga$, $\ph$ and $\th$. They are equivalent to Eqs. (\ref{6.2.19}) and (\ref{6.2.20}), which are the rules for symplectic matrices. Not only is the preceding analysis an interesting application of Schmidt decompositions, but it is also a constructive proof that Sp(2) and SO(1,2) are isomorphic (because the additional input and output rotations are isomorphic).

\subsec{6.3 SU(2) and SO(3)}

In Secs. 5.2 and 5.3, we stated that Schmidt decompositions of unitary and orthogonal matrices are not interesting, because these matrices are their own decompositions. In this section, we compare decompositions of the form $M = M_3M_2M_1$, where $M_i$ is a fundamental matrix, because such decompositions are useful for rotation matrices.

Every member of SU(2) can be written in the form
\be U = \left[\begin{array}{cc} c_0 + in_1s_0 & (in_2 - n_3)s_0 \\ (in_2 + n_3)s_0 & c_0 - in_1s_0 \end{array}\right], \label{6.3.1} \ee
where $c_0 = \cos(k)$, $s_0 = \sin(k)$, $k = (k_1^2 + k_2^2 + k_3^2)^{1/2}$ and $n_i = k_i/k$. For SU(2), the fundamental matrices
\be U_1 = \left[\begin{array}{cc} e_1 & 0 \\ 0 & e_1^* \end{array}\right] , \ \ 
U_2 = \left[\begin{array}{cc} c_2 &is_2 \\is_2 & c_2 \end{array}\right], \ \ 
U_3 = \left[\begin{array}{cc} c_3 & -s_3 \\ s_3 & c_3 \end{array}\right], \label{6.3.2} \ee
where $c_i = \cos(l_i)$, $s_i = \sin(l_i)$ and $e_1 = \exp(il_1) = c_1 + is_1$. Their triple product
\ba U_3U_2U_1 &= &\left[\begin{array}{cc} c_3 & -s_3 \\ s_3 & c_3 \end{array}\right]
\left[\begin{array}{cc} c_2 &is_2 \\is_2 & c_2 \end{array}\right]
\left[\begin{array}{cc} e_1 & 0 \\ 0 & e_1^* \end{array}\right] \nonumber \\
&= &\left[\begin{array}{cc} c_3 & -s_3 \\ s_3 & c_3 \end{array}\right]
\left[\begin{array}{cc} c_2e_1 &is_2e_1^* \\is_2e_1 & c_2e_1^* \end{array}\right] \nonumber \\
&= &\left[\begin{array}{cc} c_3c_2e_1 - is_3s_2e_1 & ic_3s_2e_1^* - s_3c_2e_1^* \\ s_3c_2e_1 + ic_3s_2e_1 & c_3c_2e_1^* + is_3s_2e_1^* \end{array}\right]. \label{6.3.3} \ea
In order for matrix (\ref{6.3.3}) to equal matrix (\ref{6.3.1}),
\ba c_0&= &c_3c_2c_1 + s_3s_2s_1, \label{6.3.4} \\
n_1s_0 &= &c_3c_2s_1 - s_3s_2c_1, \label{6.3.5} \\
n_2s_0 &= &c_3s_2c_1 + s_3c_2s_1 \label{6.3.6} \\
n_3s_0 &= &s_3c_2c_1 - c_3s_2s_1. \label{6.3.7} \ea
Equations (\ref{6.3.4}) -- (\ref{6.3.7}) were obtained by comparing the real and imaginary parts of the matrices.
Notice that they  specify the coefficients $k_i$ as functions of $l_j$. It is easy to verify that $c_0^2 + (n_1s_0)^2 + (n_2s_0)^2 + (n_3s_0)^2 = 1$, as it should do.

Likewise, every matrix in SO(3) can be written in the form
\be R = \left[\begin{array}{ccc} c_0 + n_1^2d_0 & -n_3s_0 + n_1n_2d_0 & n_2s_0 + n_1n_3d_0 \\ n_3s_0 + n_2n_1d_0 & c_0 + n_2^2d_0 & -n_1s_0 + n_2n_3d_0 \\ -n_2s_0 + n_3n_1d_0 & n_1s_0 + n_3n_2d_0 & c_0 + n_3^2d_0 \end{array}\right], \label{6.3.11} \ee
where $c_0 = \cos(k)$, $d_0 = 1 - c_0$, $s_0 = \sin(k)$, $k = (k_1^2 + k_2^2 + k_3^2)^{1/2}$ and $n_i = k_i/k$. For SO(3), the fundamental matrices
\be R_1 = \left[\begin{array}{ccc} 1 & 0 & 0 \\ 0 & c_1 & -s_1 \\ 0 & s_1 & c_1 \end{array}\right], \ \ 
R_2 = \left[\begin{array}{ccc} c_2 & 0 & s_2 \\ 0 & 1 & 0 \\ -s_2 & 0 & c_2 \end{array}\right], \ \ 
R_3 = \left[\begin{array}{ccc} c_3 & -s_3 & 0 \\ s_3 & c_3 & 0 \\ 0 & 0 & 1 \end{array}\right], \label{6.3.12} \ee
where $c_i = \cos(l_i)$ and $s_i = \sin(l_i)$. 
Their triple product
\be R_3R_2R_1 = \left[\begin{array}{ccc} c_3c_2 & c_3s_2s_1 - s_3c_1 & c_3s_2c_1 + s_3s_1 \\
s_3c_2 & s_3s_2s_1 + c_3c_1 & s_3s_2c_1 - c_3s_1 \\ -s_2 & c_2s_1 & c_2c_1 \end{array}\right]. \label{6.3.13} \ee
In order for matrix (\ref{6.3.13}) to equal matrix (\ref{6.3.11}),
\ba 2c_0 + 1 &= &s_3s_2s_1 + c_3c_2 + c_3c_1 + c_2c_1, \label{6.3.14} \\
2n_1s_0 &= &c_2s_1 + c_3s_1 - s_3s_2c_1,  \label{6.3.15} \\
2n_2s_0 &= &c_3s_2c_1 + s_3s_1 + s_2, \label{6.3.16} \\ 
2n_3s_0 &= &s_3c_2 + s_3c_1 - c_3s_2s_1.  \label{6.3.17} \ea
Equation (\ref{6.3.14}) was obtained by comparing the traces of the matrices, whereas Eqs. (\ref{6.3.15}) -- (\ref{6.3.17}) were obtained by comparing the differences of pairs of elements. Notice that they define the coefficients $k_i$ as functions of $l_j$. It is tedious, but straightforward, to verify that $c_0^2 + (n_1s_0)^2 + (n_2s_0)^2 + (n_3s_0)^2 = 1$, as it should do.

At first glance, Eqs. (\ref{6.3.14}) -- (\ref{6.3.17}) look nothing like their counterparts, Eqs. (\ref{6.3.4}) -- (\ref{6.3.7}), respectively. However, matrix (\ref{6.3.11}) is the exponential of canonical generators, whose commutation relations have coefficients of 1 on their right sides. In contrast, matrix (\ref{6.3.1}) is the exponential of generators, whose commutation relations have coefficients of 2 on their right sides. In order to make a fair comparison between the results, one must replace the SU(2) coefficients $k_i$ and $l_i$ by $k_i/2$ and $l_i/2$, respectively.

The half-coefficient equations (\ref{6.3.5}) -- (\ref{6.3.7}) must be rewritten in terms of full coefficients. By using the identities
\ba c_2^2c_1^2 - s_2^2s_1^2 &= &(c_2^2 - s_2^2 + c_1^2 - s_1^2)/2, \label{6.3.21} \\
c_2^2c_1^2 + s_2^2s_1^2 &= &[(c_2^2 - s_2^2)(c_1^2 - s_1^2) + 1]/2, \label{6.3.22} \ea
one can show that
\ba 4n_1s_0c_0 &= &4(c_3c_2s_1 - s_3s_2c_1)(c_3c_2c_1 + s_3s_2s_1) \nonumber \\
&= &4s_1c_1(c_3^2c_2^2 - s_3^2s_2^2) - 4s_3c_3s_2c_2(c_1^2 - s_1^2) \nonumber \\
&= &(2s_1c_1)(c_3^2 - s_3^2 + c_2^2 - s_2^2) - (2s_3c_3)(2s_2c_2)(c_1^2 - s_1^2), \label{6.3.23} \\
4n_2s_0c_0 &= &4(c_3s_2c_1 + s_3c_2s_1)(c_3c_2c_1 + s_3s_2s_1) \nonumber \\
&= &4s_2c_2(c_3^2c_1^2 + s_3^2s_1^2) + 4s_3c_3s_1c_1(c_2^2 + s_2^2) \nonumber \\
&= &(2s_2c_2)[(c_3^2 - s_3^2)(c_1^2 - s_1^2) + 1] + (2s_3c_3)(2s_1c_1), \label{6.3.24} \\
4n_3s_0c_0 &= &4(s_3c_2c_1 - c_3s_2s_1)(c_3c_2c_1 + s_3s_2s_1) \nonumber \\
&= &4s_3c_3(c_2^2c_1^2 - s_2^2s_1^2) - 4s_2c_2s_1c_1(c_3^2 - s_3^2)\nonumber \\
&= &(2s_3c_3)(c_2^2 - s_2^2 + c_1^2 - s_1^2) - (2s_2c_2)(2s_1c_1)(c_3^2 - s_3^2). \label{6.3.25} \ea
Equations (\ref{6.3.23}) -- (\ref{6.3.25}) are equivalent to Eqs. (\ref{6.3.15}) -- (\ref{6.3.17}), respectively. Provided that one uses the associated fundamental matrices in SU(2) and SO(3), the decomposition formulas are identical.
Consequently, one does not need to solve Eqs. (\ref{6.3.5}) -- (\ref{6.3.7}) explicitly for the $l_j/2$ coefficients as functions of $k_i/2$. One can use Eqs. (\ref{5.3.3b}) to determine the full coefficients $l_j$, then divide the results by 2.

There is a formalism, called the Jones--Stokes formalism, which illustrates the relations between SU(2) and SO(3), and facilitates the derivations of useful results.
In the notation of \cite{gor00}, for every (complex) Jones vector $|s\> = [u, v]^t$, there exists an associated (real) Stokes vector $\vs = [s_1, s_2, s_3]^t$. Let $\vsi = [\si_1, \si_2, \si_3]^t$ be the vector whose components are the spin matrices (\ref{2.2.6}). Then the Stokes vector $\vs = \<s|\vsi|s\>$. In the language of quantum mechanics, $|s\>$ is the state vector and each component of the Stokes vector is the expectation value of the corresponding spin matrix (operator). Written explicitly,
\ba  s_1 &= &[u^*, v^*][u, -v]^t \ = \ |u|^2 - |v|^2, \label{6.3.31} \\
s_2 &= &[u^*, v^*][v, u]^t \ = \ u^*v + v^*u, \label{6.3.32} \\
s_3 &= &[u^*, v^*][-iv, iu]^t \ = \ (u^*v - v^*u)/i. \label{6.3.33} \ea
SU(2) matrix operations in Jones space preserve the norm $\<s|s\> = |u|^2 + |v|^2$, whereas SO(3) operations in Stokes space preserve the norm $\vs\cdot\vs = s_1^2 + s_2^2 + s_3^2 = (|u|^2 + |v|^2)^2$.

Let $U$ be a unitary matrix and $R$ be a rotation matrix, and let $|s\>$ and $\vs$\, be the input Jones and Stokes vectors, respectively. Then the transformed (output) Jones vector $|s'\> = U|s\>$, from which it follows that the output Stokes vector
\be \vs\,' = \<s'|\vsi|s'\> = \<s|U^\d \vsi U|s\>. \label{6.3.34} \ee
Alternatively, one can write the output Stokes vector,
\be \vs\,' = R\vs = R\<s|\vsi|s\> = \<s|R\vsi|s\>, \label{6.3.35} \ee
as a rotated version of the input vector.
In Eq. (\ref{6.3.35}), the last step is possible because a linear combination of expectation values equals the expectation value of the same linear combination of operators. Equations (\ref{6.3.34}) and (\ref{6.3.35}) are true for arbitrary input vectors, so it must also be true that
\be R\vsi = U^\d\vsi U. \label{6.3.36} \ee
Equation (\ref{6.3.36}) is the operational definition of the rotation matrix $R$ associated with the unitary matrix $U$.

The fundamental matrices of SU(2) were stated in Eqs. (\ref{6.3.2}), which were based on generators (\ref{2.2.4}).
In the Jones--Stokes formalism, which is based on generators (\ref{2.2.6}), the signs of the $s$ terms in the third matrix are changed. Under transformation 1, $u' = eu = (c + is)u$ and $v' = e^*v = (c - is)v$, from which it follows that
\ba s_1' &= &|u|^2 - |v|^2, \\
s_2' &= &(d - it)u^*v + (d + it)v^*u \nonumber \\
&= &d(u^*v + v^*u) + t(u^*v - v^*u)/i, \\
s_3' &= &[(d - it)u^*v - (d + it)v^*u]/i \nonumber\\
&= &d(u^*v - v^*u)/i - t(u^*v + v^*u), \label{6.3.41} \ea
where $d = c^2 - s^2$ and $t = 2cs$. (In the remainder of this section, $d \neq1 - c$.)
Under transformation 2, $u' = cu + isv$ and $v' = cv + isu$, from which it follows that
\ba s_1' &= &(cu^* - isv^*)(cu + isv) - (cv^* - isu^*)(cv + isu) \nonumber \\
&= &c^2|u|^2 + ics(u^*v - v^*u) + s^2|v|^2 \nonumber \\
&&-(c^2|v|^2 - ics(u^*v - v^*u) + s^2|u|^2) \nonumber \\
&= &d(|u|^2 - |v|^2) - t(u^*v - v^*u)/i, \\
s_2' &= &(cu^* - isv^*)(cv + isu) + (cv^* - isu^*)(cu + isv) \nonumber \\
&= &c^2u^*v + ics(|u|^2 - |v|^2) + s^2v^*u \nonumber \\
&&+\ c^2v^*u - ics(|u|^2 - |v|^2) + s^2u^*v \nonumber \\
&= &(u^*v + v^*u), \\
s_3' &= &\{c^2u^*v + ics(|u|^2 - |v|^2) + s^2v^*u \nonumber\\
&&-\ [c^2v^*u - ics(|u|^2 - |v|^2) + s^2u^*v]\}/i \nonumber \\
&= &d(u^*v - v^*u)/i + t(|u|^2 - |v|^2). \label{6.3.42} \ea
Under transformation 3, $u' = cu + sv$ and $v' = cv - su$, from which it follows that
\ba s_1' &= &(cu^* + sv^*)(cu + sv) - (cv^* - su^*)(cv - su) \nonumber \\
&= &c^2|u|^2 + cs(u^*v + v^*u) + s^2|v|^2 \nonumber \\
&&- [c^2|v|^2 - cs(u^*v + v^*u) + s^2|u|^2] \nonumber \\
&= &d(|u|^2 - |v|^2) + t(u^*v + v^*u), \\
s_2' &= &(cu^* + sv^*)(cv - su) + (cv^* - su^*)(cu + sv) \nonumber \\
&= &c^2u^*v - cs(|u|^2 - |v|^2) - s^2v^*u \nonumber \\
&&+\ c^2v^*u - cs(|u|^2 - |v|^2) - s^2u^*v \nonumber \\
&= &d(u^*v + v^*u) - t(|u|^2 - |v|^2), \\
s_3' &= &\{c^2u^*v - cs(|u|^2 - |v|^2) - s^2v^*u \nonumber \\
&&-\ [c^2v^*u - cs(|u|^2 - |v|^2) - s^2u^*v]\}/i \nonumber \\
&= &(u^*v - v^*u)/i. \label{6.3.43} \ea
It is easy to verify that all three transformations preserve $|u|^2 + |v|^2$, as stated above.
By writing the preceding results in the matrix form $\vs\,' = R\vs$, one obtains the fundamental rotation matrices
\be R_1 = \left[\begin{array}{ccc} 1 & 0 & 0 \\ 0 & d & t \\ 0 & -t & d \end{array}\right], \ \ 
R_2 = \left[\begin{array}{ccc} d & 0 & -t \\ 0 & 1 & 0 \\ t & 0 & d \end{array}\right], \ \ 
R_3 = \left[\begin{array}{ccc} d & t & 0 \\ -t & d & 0 \\ 0 & 0 & 1 \end{array}\right]. \label{6.3.44} \ee
Each matrix $R_i$ represent a passive rotation about the $i$ axis, which is easy to visualize.

One can obtain equivalent results by multiplying the spin and unitary matrices [Eq. (\ref{6.3.36})]. It is easy to verify that
\ba U_1^\d \si_1U_1 = \si_1, \ \ U_1^\d \si_2U_1 = d\si_2 + t\si_3, \ \ U_1^\d \si_3U_1 = d\si_3 - t\si_2, \label{6.3.45} \\
U_2^\d \si_1U_2 = d\si_1 - t\si_3, \ \ U_2^\d \si_2U_2 = \si_2, \ \ U_2^\d \si_3U_2 = d\si_3 + t\si_1, \label{6.3.46} \\
U_3^\d \si_1U_3 = d\si_1 + t\si_2, \ \ U_3^\d \si_2U_3 = d\si_2 - t\si_1, \ \ U_3^\d \si_3U_3 = \si_3. \label{6.3.47} \ea
By defining $\vsi' = U^\d\vsi U$ and rewriting the preceding results in the matrix form $\vsi' = R\vsi$, one obtains the rotation matrices (\ref{6.3.44}).
In the first approach, the Jones vector $|s\>$ is transformed and used to evaluate the expectation values of the spin operators, whereas in the second, the input Jones vector is used to evaluate the expectation values of the transformed spin operators. These approaches correspond to the Schr\"odinger and Heisenberg pictures of quantum mechanics, respectively.

We prefer active transformations to passive ones, so in the rest of this section, we will use the fundamental matrices $U_i = \exp(-i\si_ik_i/2)$, as did the authors of \cite{gor00}. Changing the signs of the exponents has the effect of changing the signs of the $s_i$ terms in Eqs. (\ref{6.3.2}) and the $t$ terms in Eqs. (\ref{6.3.44}). With these changes, the rotation matrices represent active rotations.

In spin-vector notation, every unitary matrix can be written in the form
\be U = c\si_0 - is\vn\cdot\vsi, \label{6.3.51} \ee
where $c$, $s$ and $\vn$ were defined after Eq. (\ref{6.3.1}), and the argument of the trigonometric functions is the generator coefficient (half angle) $k/2$.
Notice that the sign of the last term in Eq. (\ref{6.3.51}) is negative.
How does a general unitary transformation affect the spin vector? It is easy to verify that
\ba U^\d\vsi U &= &(c\si_0 + is\vn\cdot\vsi)\vsi(c\si_0 - is\vn\cdot\vsi) \nonumber \\
&= &c^2\vsi + ics[(\vn\cdot\vsi)\vsi - \vsi(\vn\cdot\vsi)] + s^2(\vn\cdot\vsi)\vsi(\vn\cdot\vsi). \label{6.3.52} \ea
The spin matrices have the properties $\si_j^2 = -\si_0$, where $\si_0$ is the identity matrix, and $\si_j\si_k$ $= \pm i\si_l$, where the plus (minus) sign applies if the indices $j$, $k$ and $l$ are in positive (negative) cyclic order. By using these properties, one can verify the spin-vector identities \cite{gor00}
\ba \vsi(\vn\cdot\vsi) &=& \vn\si_0 + i\vn\times\vsi, \label{6.3.53} \\
(\vn\cdot\vsi)\vsi &= &\vn\si_0 - i\vn\times\vsi, \label{6.3.54} \\
(\vm\cdot\vsi)(\vn\cdot\vsi) &= &(\vm\cdot\vn)\si_0 + i(\vm\times\vn)\cdot\vsi, \label{6.3.55} \\
(\vn\cdot\vsi)\vsi(\vn\cdot\vsi) &= &2\vn(\vn\cdot\vsi) - n^2\vsi. \label{6.3.56} \ea
By using identities (\ref{6.3.53}), (\ref{6.3.54}) and (\ref{6.3.56}), one finds that
\ba U^\d\vsi U &= &c^2\vsi + 2cs\vn\times\vsi + s^2[2\vn(\vn\cdot\vsi) - \vsi] \nonumber \\
&= &(c^2 - s^2)\vsi + 2cs\vn\times\vsi + (2s^2)\vn(\vn\cdot\vsi) \nonumber \\
&= &[d\si_0 + t\vn\times +\ (1 - d)\vn\vn\cdot]\vsi, \label{6.3.57} \ea
where, as stated above, $d = c^2 - s^2$ and $t = 2cs$. By comparing Eqs. (\ref{6.3.36}) and (\ref{6.3.57}), one finds that the rotation matrix
\be R = d\si_0 + t\vn\times +\ (1 - d)\vn\vn\cdot. \label{6.3.58} \ee
Equation  (\ref{6.3.58}) is equivalent to Eq. (\ref{d1}), so definition (\ref{6.3.36}) produces the canonical form of the matrix.
In words, every unitary transformation in Jones space, which is specified by the generator parameters $k/2$ and $\vn$, corresponds to a rotation in Stokes space about the axis $\vn$ through the angle $k$.
Henceforth, we will refer to $\vn$ and $k$ as the axis (direction) vector and rotation full-angle, respectively.

Equation (\ref{6.3.36}) defines the rotation matrix $R$ associated with the unitary matrix $U$. How are matrix products related? By multiplying the equation $R_2\vsi = U_2^\d\vsi U_2$ by $U_1^\d$ on the left and $U_1$ on the right, one finds that
\be U_1^\d(R_2\vsi)U_1 = U_1^\d(U_2^\d\vsi U_2)U_1 = (U_2U_1)^\d\vsi(U_2U_1). \label{6.3.59a}\ee
One also finds that
\be U_1^\d(R_2\vsi)U_1 = R_2(U_1^\d\vsi U_1) = (R_2R_1)\vsi, \label{6.3.59b}\ee
where the first step is possible because the transformation of a linear combination of spin matrices equals the same linear combination of transformed matrices. By comparing Eqs. (\ref{6.3.59a}) and (\ref{6.3.59b}), one finds that
\be (R_2R_1)\vsi = (U_2U_1)^\d \vsi (U_2U_1). \label{6.3.60} \ee
In words, the rotation matrix associated with the unitary product $U_2U_1$ is the orthogonal product $R_2R_1$.
This result shows that the rules of multiplication are preserved, so the groups SU(2) and SO(3) are isomorphic.

Equation (\ref{6.3.60}) is deceptively simple. To illustrate its importance, we will calculate the products of unitary and orthogonal matrices directly. Let $U_3 = U_2U_1$ be a unitary product matrix. Then it follows from Eq. (\ref{6.3.51}) that
\ba U_3 &= &(c_2\si_0 - is_2\vn_2\cdot\vsi)(c_1\si_0 - is_1\vn_1\cdot\vsi) \nonumber \\
&= &c_2c_1\si_0 - ic_2s_1\vn_1\cdot\vsi - ic_1s_2\vn_2\cdot\vsi - s_2s_1(\vn_2\cdot\vsi)(\vn_1\cdot\vsi). \label{6.3.61} \ea
By using identity (\ref{6.3.55}), one finds that
\be U_3 = (c_2c_1 - s_2s_1\vn_2\cdot\vn_1)\si_0 - i(c_2s_1\vn_1 + c_1s_2\vn_2 + s_2s_1\vn_2\times\vn_1)\cdot\vsi. \label{6.3.62} \ee
One can rewrite Eq. (\ref{6.3.62}) in the form of Eq. (\ref{6.3.51}) by defining the scalar and vector quantities
\ba c_3 &= &c_2c_1 - s_2s_1\vn_2\cdot\vn_1, \label{6.3.63} \\
s_3\vn_3 &= &c_2s_1\vn_1 + c_1s_2\vn_2 + s_2s_1\vn_2\times\vn_1, \label{6.3.64} \ea
respectively. It is easy to verify that $c_3^2 + (s_3\vn_3)\cdot(s_3\vn_3) = 1$, as it should do. Equations (\ref{6.3.63}) and (\ref{6.3.64}) are the multiplication rules for SU(2), written in terms of direction vectors and half angles ($\vn$ and $k/2$). For the special case in which $\vn_2 = \vn_1$, $c_3 = c_2c_1 - s_2s_1$ and $s_3 = s_2c_1 + c_2s_1$, whereas for the complementary case in which $\vn_2 = -\vn_1$, $c_3 = c_2c_1 + s_2s_1$ and $s_3 = s_1c_2 - c_1s_2$. The half angle of the product matrix is the sum (difference) of the half angles of the constituent matrices. The preceding calculation is simple, but the following one is not.

Let $R_3 = R_2R_1$ be an orthogonal product matrix. Then
\ba R_3 &= &[d_2 + t_2\vn_2\times + \ (1 - d_2)\vn_2\vn_2\cdot]
[d_1 + t_1\vn_1\times + \ (1 - d_1)\vn_1\vn_1\cdot] \nonumber \\
&= &d_2d_1 + d_2t_1\vn_1\times + \ d_2(1 - d_1)\vn_1\vn_1\cdot \label{6.3.71} \\
&&+\ d_1t_2\vn_2\times + \ t_2t_1\vn_2\times\vn_1\times + \ t_2(1 - d_1)(\vn_2\times\vn_1)\vn_1\cdot \nonumber \\
&&+\ (1 - d_2)d_1\vn_2\vn_2\cdot + \ (1 - d_2)t_1(\vn_2\vn_2\cdot)\vn_1\times + \ (1 - d_2)(1 - d_1)(\vn_2\cdot\vn_1)\vn_2\vn_1\cdot. \nonumber \ea
By considering the effects of the operators on an arbitrary vector, one can verify the vector identities
\ba \vn_2\cdot\vn_1\times &= &(\vn_2\times\vn_1)\cdot, \label{6.3.72} \\
\vn_2\times\vn_1\times &= &-(\vn_2\cdot\vn_1) + \vn_1\vn_2\cdot, \label{6.3.73} \\
(\vn_2\times\vn_1)\times &= &\vn_1\vn_2\cdot - \ \vn_2\vn_1\cdot. \label{6.3.74} \ea
%
By using identity (\ref{6.3.72}), one can rewrite the eighth term in Eq. (\ref{6.3.71}) as $(1 - d_2)t_1\vn_2(\vn_2\times\vn_1)\cdot$. Then, after some regrouping, one finds that
\ba R_3 &= &d_2d_1 + d_2t_1\vn_1\times +\ d_1t_2\vn_2\times + \ t_2t_1\vn_2\times\vn_1\times \nonumber \\
&&+\ t_2(1 - d_1)(\vn_2\times\vn_1)\vn_1\cdot + \ (1 - d_2)t_1\vn_2(\vn_2\times\vn_1)\cdot \nonumber \\
&&+\ d_2(1 - d_1)\vn_1\vn_1\cdot + \ (1 - d_2)d_1\vn_2\vn_2\cdot + \ (1 - d_2)(1 - d_1)\de\vn_2\vn_1\cdot, \label{6.3.75}\ea
where, once again, the dot product $\de = \vn_2\cdot\vn_1$.
For the special cases in which $\vn_2 = \pm\vn_1$, 
\be R_3 = (d_2d_1 \mp t_2t_1) + (t_1d_2 \pm d_2t_2)\vn\times +\ (1 - d_2d_1 \pm t_2t_1)\vn\vn\cdot. \label{6.3.76}\ee
The full angle of the product matrix is the sum (difference) of the full angles of the constituent matrices.

With $R_3$ known, one can extract formulas for $d_3 = [\tr(R_3) - 1]/2$ and $t_3\vn_3\times = (R_3 - R_3^t)/2$ [Eq. (\ref{6.3.58})].
It is easy to verify the trace identities
\be \tr(\vn_i\times) = 0, \ \ 
\tr(\vn_1\vn_2\cdot) = \vn_2\cdot\vn_1, \ \ 
\tr(\vn_2\times\vn_1\times) = -2\de. \label{6.3.77} \ee
By combining Eqs. (\ref{6.3.75}) and (\ref{6.3.77}), one finds that
\ba \tr(R_3) &= &3d_2d_1 + 0 + 0 - 2t_2t_1\de + 0 + 0 \nonumber \\
&&+\ d_2(1 - d_1) + (1 - d_2)d_1 + (1 - d_2)(1 - d_1)\de^2 \nonumber \\
&= &3d_2d_1 - 2t_2t_1\de + (1 - d_2)(1 - d_1)(\de^2 - 1) \nonumber \\
&&+\ d_2(1 - d_1) + (1 - d_2)d_1 + (1 - d_2)(1 - d_1) \nonumber \\
&= &2d_2d_1 - 2t_2t_1\de - (1 - d_2)(1 - d_1)(1 - \de^2) + 1. \label{6.3.78} \ea
By comparing Eq. (\ref{6.3.78}) to the aforementioned trace formula, one finds that 
\ba d_3 &= &d_2d_1 - t_2t_1\de - (1 - d_2)(1 - d_1)(1 - \de^2)/2 \nonumber \\
&= &(1 + d_2)(1 + d_1)/2 - t_2t_1\de + (1 - d_2)(1 - d_1)\de^2/2 - 1. \label{6.3.79} \ea

Three of the terms in Eq. (\ref{6.3.75}) are symmetric and do not contribute to $R - R^t$. By retaining only the nonsymmetric terms, one finds that
\ba R_3 &\approx &d_2t_1\vn_1\times +\ d_1t_2\vn_2\times + \ t_2t_1\vn_2\times\vn_1\times \nonumber \\
&&+\ t_2(1 - d_1)(\vn_2\times\vn_1)\vn_1\cdot + \ (1 - d_2)t_1\vn_2(\vn_2\times\vn_1)\cdot \nonumber \\
&&+\ (1 - d_2)(1 - d_1)\de\vn_2\vn_1\cdot. \label{6.3.80}\ea
The transpose $R_3^t = R_1^tR_2^t$, where $R_i^t(s) = R_i(-s)$. Hence, one transposes $R_3$ by exchanging the subscripts 1 and 2, and changing the signs of $s_i$. For the first and second terms,
\be R_3 - R_3^t = 2d_2t_1\vn_1\times +\ 2d_1t_2\vn_2\times. \label{6.3.81} \ee
For the third term,
\ba R_3 - R_3^t &= &t_2t_1(\vn_2\times\vn_1\times -\ \vn_1\times\vn_2\times) \nonumber \\
&= &t_2t_1(\vn_1\vn_2\cdot -\ \vn_2\vn_1\cdot) \nonumber \\
&= &t_2t_1(\vn_2\times\vn_1)\times. \label{6.3.82} \ea
[Identities (\ref{6.3.73}) and (\ref{6.3.74}) were used.]
For the fourth and fifth terms,
\ba R_3 - R_3^t &= &t_2(1 - d_1)(\vn_2\times\vn_1)\vn_1\cdot + \ t_1(1 - d_2)\vn_2(\vn_2\times\vn_1)\cdot \nonumber \\
&&+\ t_1(1 - d_2)(\vn_1\times\vn_2)\vn_2\cdot + \ t_2(1 - d_1)\vn_1(\vn_1\times\vn_2)\cdot \nonumber \\
&= &t_2(1 - d_1)[(\vn_2\times\vn_1)\vn_1\cdot -\ \vn_1(\vn_2\times\vn_1)\cdot] \nonumber \\
&&+\ t_1(1 - d_2)[\vn_2(\vn_2\times\vn_1)\cdot -\ (\vn_2\times\vn_1)\vn_2\cdot] \nonumber \\
&= &t_2(1 - d_1)(\vn_2 - \de\vn_1)\times +\ t_1(1 - d_2)(\vn_1 - \de\vn_2)\times \nonumber \\
&= &[(1 - d_2)t_1\vn_1 + (1 - d_1)t_2\vn_2]\times -\ \de[t_2(1 - d_1)\vn_1 +t_1(1 - d_2)\vn_2]\times. \label{6.3.83} \ea
[Identity (\ref{6.3.74}) was used.]
For the sixth term,
\ba R_3 - R_3^t &= &(1 - d_2)(1 - d_1)\de(\vn_2\vn_1\cdot -\ \vn_1\vn_2\cdot) \nonumber \\
&= &-(1 - d_2)(1 - d_1)\de(\vn_2\times\vn_1)\times. \label{6.3.84} \ea
[Identity (\ref{6.3.74}) was used.]
By adding the preceding contributions to $R_3 - R_3^t$ and comparing the result to the aforementioned cross-product formula, one finds that
\ba 2t_3\vn_3 &= &[2d_2t_1 + t_1(1 - d_2) - t_2(1 - d_1)\de]\vn_1 \nonumber \\
&&+\ [2d_1t_2 + t_2(1 - d_1) - t_1(1 - d_2)\de]\vn_2 \nonumber \\
&&+\ [t_2t_1 - (1 - d_2)(1 - d_1)\de](\vn_2\times\vn_1). \label{6.3.85} \ea
Equations (\ref{6.3.79}) and (\ref{6.3.85}) are the multiplication rules for SO(3), written in terms of direction vectors and full angles ($\vn$ and $k$). They are unilluminating.

Equations (\ref{6.3.79}) and (\ref{6.3.85}) involve full angles, whereas Eqs. (\ref{6.3.63}) and (\ref{6.3.64}) involve half angles. By using trigonometric identities, one can rewrite the second line of Eq. (\ref{6.3.79}) as
\be d_3 = 2(c_2c_1 - s_2s_1\de)^2 - 1. \label{6.3.91} \ee
Equation (\ref{6.3.91}) is consistent with Eq. (\ref{6.3.63}).
It is easy to verify that
\ba t_1(1 + d_2) - t_2(1 - d_1)\de &= &4(c_2c_1 - s_2s_1\de)c_2s_1, \label{6.3.92} \\
t_2(1 + d_1) - t_1(1 - d_2)\de &= &4(c_2c_1 - s_2s_1\de)c_1s_2, \label{6.3.93} \\
t_2t_1 - (1 - d_2)(1 - d_1)\de &= &4(c_2c_1 - s_2s_1\de)s_2s_1. \label{6.3.94} \ea
By combining Eq. (\ref{6.3.85}) with Eqs. (\ref{6.3.92}) -- (\ref{6.3.94}) and dividing the result by 2, one finds that
\be t_3\vn_3 = 2(c_2c_1 - s_2s_1\de)[c_2s_1\vn_1 + c_1s_2\vn_2 +s_2s_1(\vn_2\times\vn_1)]. \label{6.3.95}\ee
Equation (\ref{6.3.95}) is consistent with Eqs. (\ref{6.3.63}) and (\ref{6.3.64}). These results verify that SO(3) is isomorphic to SU(2).

A related analysis was provived by the authors of \cite{rui13}, who worked with half angles throughout. [See the second line of Eq. (\ref{6.3.57}).] Not only did they derive expressions~for $\tr(R) = 4c^2 - 1$ and $(R - R^t)/2 = 2cs\vn\times$, but they also derived an expression for $(R + R^t)/2 = (2c^2 - 1)I + 2s^2\vn\vn\cdot$. (They referred to $c$, $sn_1$, $sn_2$ and $sn_3$ as quaternion components, because $c\si_0 - is\vn\cdot\vsi$ is a matrix representation of a quaternion.)

\newpage

\sec{7. Summary}

In this tutorial, we discussed the properties of the matrix groups Sp(2), SU(2), SU(1,1), SO(3) and SO(1,2), which arise in Hamiltonian dynamics and optics, frequency conversion, parametric amplification, rotation in three space dimensions, and Lorentz transformation in time and two space dimensions, respectively. The symplectic group Sp(2) consists of real $2 \times 2$ matrices, the special unitary groups SU(2) and SU(1,1) consist of complex $2 \times 2$ matrices, and the special orthogonal groups SO(3) and SO(1,2) consist of real $3 \times 3$ matrices.

In Sec. 2, we stated the canonical forms of the aforementioned matrices [Eqs. (\ref{2.1.3}),~(\ref{2.2.2}), (\ref{2.2.12}), (\ref{2.3.2}) and (\ref{2.3.12})], which are defined by equations of the form $M^\d SM = S$, where $S$ is a real structure matrix. Every matrix $M$ can be written as the exponential of a generating matrix $G$. The generating matrices are defined by equations of the form $G^\d S + SG = 0$. Both sets of equations are summarized in Tab.~1. By considering them, we showed that the matrices all have three free parameters, the physical significances of which vary from group to group. We also showed that every generating matrix can be written as the linear combination $G = G_ik_i$, where $G_i$ is a basis generator [Eqs. (\ref{2.1.6}), (\ref{2.2.4}), (\ref{2.2.14}), (\ref{2.3.4}) and (\ref{2.3.14})], $k_i$ is a real generator coefficient and repeated indices imply summation. The generating matrices also have three free parameters (the generator coefficients), which can vary continuously. Hence, the matrices are members of Lie groups, whereas the generating matrices are members of the associated Lie algebras.
The generators of Sp(2), SU(1,1) and SO(1,2) satisfy equivalent commutation relations [Eqs. (\ref{2.1.7}), (\ref{2.2.15}) and (\ref{2.3.15})], as do the generators of SU(2) and SO(3) [Eqs. (\ref{2.2.5}) and (\ref{2.3.5})].

In Sec. 3, we used the Cayley--Hamilton theorem to exponentiate the $2 \times 2$ generators, to obtain the generator forms of symplectic, unitary and indefinite unitary matrices [Eqs. (\ref{3.1.4}), (\ref{3.2.4}) and (\ref{3.3.4})]. These forms are consistent with the canonical forms stated in Sec. 2 [Eqs. (\ref{2.1.3}), (\ref{2.2.2}) and (\ref{2.2.12})].
In Sec. 4, we exponentiated the $3 \times 3$ generators, to obtain the generator forms of orthogonal and indefinite orthogonal matrices [Eqs. (\ref{4.1.7}) and (\ref{4.2.4})]. In the first case, exponentiation produced the canonical form directly [Eq. (\ref{2.3.2})], whereas in the second, it did not [Eq. (\ref{2.3.12})]. Further work was required to show that the generator and canonical forms are equivalent.
Not only does exponention produce the generator form of the matrix $M= \exp(G)$, it also produces the generator form of the inverse matrix $M^{-1} = \exp(-G)$. One can deduce the formula for the inverse matrix from the formula for the matrix by changing the signs of the odd-order terms in the Taylor expansion of the exponential function, which amounts to changing the sign of the sine (or hyperbolic sine) function in the formula.

Every real matrix has the Schmidt decomposition $M = QDP^t$, where $D$ is diagonal, and $P$ and $Q$ are orthogonal, and every complex matrix has the decomposition $M = VDU^\d$, where $U$ and $V$ are unitary. In Sec. 5, we derived Schmidt decompositions for symplectic, indefinite unitary and indefinite orthogonal matrices [Eqs. (\ref{5.1.2}) and (\ref{5.1.4}), Eqs. (\ref{5.2.3}) and (\ref{5.2.5}), and Eqs. (\ref{5.4.2}) and (\ref{5.4.3})].
Each decomposed matrix is specified by one dilation parameter ($\la$) and two angle parameters ($\th_1$ and $\th_2$).
For Sp(2), $\th_1$ and $\th_2$ are input- and output-rotation angles.
For SU(1,1), $\la$ is related to the amplification parameter, and $\th_1$ and $\th_2$ are input- and output-phase angles.
For SO(1,2), $\la$ is related to the boost (energy) parameter, and $\th_1$ and $\th_2$ are input- and output-rotation angles.
Not only are Schmidt decompositions mathematically useful, but they are also physically meaningful.
The aforementioned equations specify the decomposition parameters in terms of the generator coefficients.
Schmidt decompositions are not relevant for orthogonal matices (because they are already orthogonal). We showed that every orthogonal matrix, which corresponds to a rotation about an arbitrary axis [Eq. (\ref{4.1.7})], can be written as the product of three simpler matrices, which correspond to rotations about the coordinate axes [Eqs. (\ref{5.3.1})]. The rotation angles were specified in terms of the generator coefficients [Eqs. (\ref{5.3.3b})].

Let $A$, $B$ and $C$ be members of group one, and $X$, $Y$ and $Z$ be members of group two. Then, in order for the groups to be isomorphic, there must be a one-to-one relationship between the members of the groups ($X$, $Y$ and $Z$ are the images of $A$, $B$ and $C$, respectively) and the rules of multiplication must be preserved ($C = BA$ in group one if and only if $Z = YX$ in group two).
The natural one-to-one relationship between the matrices is based on their generator forms: $A = \exp(G_ik_i)$ and $X = \exp(H_ik_i)$, where $G_i$ and $H_i$ are generators of groups one and two, respectively. Related matrices have the same generator coefficients, but different generators.
In Sec. 6, we explained why groups whose generators have equivalent commutation relations have the same multiplication rules and, hence, are isomorphic. 
This result is well known and often used. (For example, one can represent quantum operators by matrices.)
What distinguishes this tutorial is the number of worked examples, which illustrate the power and usefulness of isomorphisms.

In Sec 6.1, we proved that Sp(2) and SU(1,1) are isomorphic by multiplying two symplectic and two indefinite unitary matrices, and showing that the product rules, written in terms of the generator coefficients and the associated hyperbolic trigonometric functions, are identical [Eqs. (\ref{6.1.4}) and (\ref{6.1.5}), and Eqs. (\ref{6.1.7}) and (\ref{6.1.8})]. Isomorphism manifests itself in Schmidt decompositions. We calculated the decompositions of both types of matrix and showed that the relations between the decomposition parameters and generator coefficients are identical [Eqs. (\ref{6.1.13}) and (\ref{6.1.15}), and Eqs. (\ref{6.1.23}) and (\ref{6.1.25})].

In Sec. 6.2, we illustrated the isomorphism between Sp(2) and SO(1,2) by calculating the Schmidt decompositions of both types of matrix. Our first decomposition of an indefinite orthogonal matrix [Eqs. (\ref{6.2.4}) and (\ref{6.2.5})] did not look like the decomposition of a symplectic matrix [Eqs. (\ref{6.1.13}) and (\ref{6.1.15})], because the generators of these matrices had different normalizations [Eqs. (\ref{2.1.7}) and (\ref{2.3.15})]. By renormalizing the generators of Sp(2), we showed that the relations between the Schmidt parameters and the generator coefficients are the same for both groups [Eqs. (\ref{6.2.6}) -- (\ref{6.2.8})].
It is better to work with generator coefficients than to work with components, because symplectic and indefinite orthogonal matrices have different sizes and the relations between them are not obvious. (A similar statement can be made about unitary and orthogonal matrices.) We also used Schmidt decompositions to study the products of two symplectic and two indefinite orthogonal matrices. By doing so, we proved that Sp(2) and SO(1,2) are indeed isomorphic [Eqs. (\ref{6.2.19}) and (\ref{6.2.20}), and Eqs. (\ref{6.2.22}) and (\ref{6.2.23})].

In Sec. 6.3, we illustrated the isomorphism between SU(2) and SO(3) by calculating triple products of both types of matrix. (Every matrix can be written as the product of three simpler matrices.) Provided that one uses generators with the same normalizations [Eqs. (\ref{2.2.5}) and (\ref{2.3.5})], the triple-product equations are identical [Eqs. (\ref{6.3.15}) -- (\ref{6.3.17}) and Eqs. (\ref{6.3.23}) -- (\ref{6.3.25})].
We also described a well-known formalism that links 
Jones space (which is two dimensional and complex) and Stokes space (which is three dimensional and real). Every unitary transformation in Jones space (which is hard to visualize) corresponds to a rotation in Stokes space (which is easy to visualize). By using this Jones--Stokes formalism, we proved that SU(2) and SO(3) are indeed isomorphic [Eqs. (\ref{6.3.59a}) -- (\ref{6.3.60})]. We also derived product rules for two unitary and two orthogonal matrices directly [Eqs. (\ref{6.3.63}) and (\ref{6.3.64}), and Eqs. (\ref{6.3.79}) and (\ref{6.3.85})], and showed that they are equivalent [Eqs. (\ref{6.3.91}) and (\ref{6.3.95})]. The complexity of the latter calculation illustrates the usefulness of the Jones--Stokes formalism.

To explain our motivations, and make the tutorial more enjoyable to read, we included several appendices, which show how the groups discussed herein arise in studies of physical systems.
In App. A, Hamiltonian dynamics is reviewed briefly. The position and momentum equations for linear systems can be written in matrix form. By examining the coefficient (generating) matrix, one finds that the evolution of a one-mode system is governed by Sp(2). A similar generating matrix arises in geometrical optics \cite{mck25a}.
Three- and four-wave interactions are reviewed briefly, in Apps. B and C, respectively. In these interactions, one or two strong pump waves drive weak signal and idler waves (sidebands). In the strong-pump, weak-sideband regime, the sideband equations are linear and can be written in matrix form. By examining the generating matrices, one finds that frequency conversion (without amplification) is governed by SU(2), whereas parametric amplification (with frequency conversion) is governed by SU(1,1).
The general forms of three-dimensional rotation and Lorentz-transformation matrices are derived from first principles in Apps. D and E, respectively.
In App. F, the Jones--Stokes formalism, which was developed to link SU(2) and SO(3), is adapted for SU(1,1) and SO(1,2). This formalism merits further study.

In summary, we described the basic properties of the Lie groups Sp(2), SU(2), SU(1,1), SO(3) and SO(1,2), and their associated Lie algebras. We also provided numerous examples of Schmidt decompositions and product rules, which illustrate the isomorphisms between Sp(2), SU(1,1) and SO(1,2), and between SU(2) and SO(3).

\newpage

\sec{Appendix A: Simple harmonic oscillator}

Symplectic transformations originate in Hamiltonian dynamics.
Let $q$ and $p$ be the displacement and momentum, respectively, of a simple harmonic oscillator. Then the oscillator dynamics are governed by the (normalized) Hamiltonian
\be H = (p^2 + q^2)/2, \label{a1} \ee
together with the Hamilton equations
\be d_t q = \pd H/\pd p, \ \ d_t p = -\pd H/\pd q, \label{a2} \ee
where $d_t$ is a time derivative. By combining Eqs. (\ref{a1}) and (\ref{a2}), one obtains the dynamical equations
\be d_t q = p, \ \ d_t p = -q. \label{a3} \ee

Now let $X = [q, p]^t = [x_1, x_2]^t$ be a variable (coordinate) vector. Then Eqs. (\ref{a3}) can be rewritten in the matrix form
\be d_t X = JX, \label{a4} \ee
where the coefficient (structure) matrix
\be J = \left[\begin{array}{cc} 0 & 1 \\ -1 & 0 \end{array}\right]. \label{a5} \ee

By applying Eqs. (\ref{a2}) to the generalized Hamiltonian
\be H = \al p^2/2 + \bt pq + \ga q^2/2, \label{a7} \ee
one obtains the generalized dynamical equations
\be d_t q = \al p + \bt q, \ \ d_t p = -\bt p - \ga q. \label{a8} \ee
Equations (\ref{a8}) also can be written in matrix form. The generalized coefficient matrix
\ba G &= &\left[\begin{array}{cc} \bt & \al \\ -\ga & -\bt \end{array}\right] \nonumber \\
&= &\bt \left[\begin{array}{cc} 1 & 0 \\ 0 & -1 \end{array}\right]
+ \al \left[\begin{array}{cc} 0 & 1 \\ 0 & 0 \end{array}\right]
+ \ga \left[\begin{array}{cc} 0 & 0 \\ -1 & 0 \end{array}\right] \nonumber \\
&= &\bt \left[\begin{array}{cc} 1 & 0 \\ 0 & -1 \end{array}\right]
+ \al' \left[\begin{array}{cc} 0 & 1 \\ 1 & 0 \end{array}\right]
+ \ga' \left[\begin{array}{cc} 0 & -1 \\ 1 & 0 \end{array}\right], \label{a9} \ea
where $\al' = (\al - \ga)/2$ and $\ga' = -(\al + \ga)/2$. Notice that the third line of Eq. (\ref{a9}) involves the generators of Sp(2), which was discussed in Sec. 2.1. The second line involves alternative generators, which produce a dilation, and horizontal and vertical shears \cite{mck25a}. They are used to study two-mode squeezing \cite{wod85,ger01}, which is the quantum analog of parametric amplification (App. B).

Although the preceding formalism only applies to real variables, a similar formalism applies to complex variables \cite{mck13}.
Let $A = (q + ip)/2^{1/2}$. Then, in the complex formulation, the Hamiltonian
\be H = \ep(A^*)^2/2 + \de|A|^2 + \ep^*A^2/2, \label{a11} \ee
where $\de$ and $\ep = \ep_r + i\ep_i$ are the frequency and coupling parameters, respectively, and the Hamilton equation
\be d_t A = -i\pd H/\pd A^*. \label{a12} \ee
By combining Eqs. (\ref{a11}) and (\ref{a12}), one obtains the amplitude equation
\be d_t A = -i\de A - i\ep A^*. \label{a13} \ee
One can reconcile Eqs. (\ref{a8}) and (\ref{a13}) by defining $\al = \de - \ep_r$, $\bt = \ep_i$ and $\ga = \de + \ep_r$, or $\de = (\al + \ga)/2$, $\ep_r = (\ga - \al)/2$ and $\ep_i = \bt$.

Equation (\ref{a13}) and its conjugate can be written in the form of Eq. (\ref{a4}), where the coefficient matrix
\ba G &= &-i\left[\begin{array}{cc} \de & \ep \\ -\ep^* & -\de \end{array}\right] \nonumber \\
&= &-\de \left[\begin{array}{cc} i & 0 \\ 0 & -i \end{array}\right]
- \ep_r \left[\begin{array}{cc} 0 & i \\ -i & 0 \end{array}\right]
+ \ep_i \left[\begin{array}{cc} 0 & 1 \\ 1 & 0 \end{array}\right]. \label{a14} \ea
Notice that the second line of Eq. (\ref{a14}) involves the generators of SU(1,1), which was discussed in Sec. 2.2.
Equations (\ref{a8}) and (\ref{a13}) are real and complex representations of the same phenomenon (oscillation), so there must be a close relation between Sp(2) and SU(1,1). This relation (isomorphism) was discussed in Sec. 6.1.

\newpage

\sec{Appendix B: Parametric amplification}

Light-wave propagation in a third-order nonlinear medium is governed by the generalized nonlinear Schr\"odinger equation (NSE)
\be \pd_z A = i\bt(i\pd_t)A + i\ga_3|A|^2A, \label{b1} \ee
where $A$ is the wave amplitude, $\pd_z = \pd/\pd z$ and $\ga_3$ is the nonlinearity (Kerr) coefficient \cite{mar08}. 
In the frequency domain, $\bt(\om) = \tsum_{n=1}^\infty \bt_n(\om_0)\om^n/n!$ is the Taylor expansion of the dispersion function about the reference frequency $\om_0$. In the time domain, the frequency difference $\om$ is replaced by the time derivative $i\pd_t$. The squared amplitude $|A|^2$ has units of power, which is proportional to the photon flux.

In degenerate four-wave mixing (FWM), which is also called modulation instability \cite{mck04}, one pump wave ($p$) interacts with signal and idler waves ($s$ and $r$), subject to the frequency-matching condition $2\om_p = \om_r + \om_s$. By substituting the three-frequency ansatz
\be A(t, z) = A_p(z)\exp(-i\om_pt) +A_r(z)\exp(-i\om_rt) + A_s(z)\exp(-i\om_st) \label{b2} \ee
in Eq. (\ref{b1}) and collecting terms of like frequency, one obtains the amplitude equations
\ba d_z A_p &= &i(\bt_p + \ga_3|A_p|^2 + 2\ga_3|A_r|^2 + 2\ga_3|A_s|^2)A_p + i2\ga_3A_p^*A_rA_s, \label{b3} \\
d_z A_r &= &i(\bt_r + 2\ga_3|A_p|^2 + \ga_3|A_r|^2 + 2\ga_3|A_s|^2)A_r + i\ga_3A_p^2A_s^*, \label{b4} \\
d_z A_s &= &i(\bt_s + 2\ga_3|A_p|^2 + 2\ga_3|A_r|^2 + \ga_3|A_s|^2)A_s + i\ga_3A_p^2A_r^*, \label{b5} \ea
where $\bt_i = \bt(\om_i)$ is a wavenumber. The factors of 1 and 2 that precede $\ga_3$ are called (non)degeneracy factors.
By combining Eqs. (\ref{b3}) -- (\ref{b5}), one finds that
\ba d_z |A_p|^2 &= &i2\ga_3(A_p^*)^2A_rA_s - i2\ga_3A_p^2A_r^*A_s^*, \label{b6} \\
d_z |A_r|^2 &= &i\ga_3A_p^2A_r^*A_s^* - i\ga_3(A_p^*)^2A_rA_s, \label{b7} \\
d_z |A_s|^2 &= &i\ga_3A_p^2A_r^*A_s^* - i\ga_3(A_p^*)^2A_rA_s, \label{b8} \ea
from which it follows that
\be d_z (|A_p|^2 + |A_r|^2 + |A_s|^2) = 0, \ \ 
d_z (|A_r|^2 - |A_s|^2) = 0. \label{b9} \ee
Equations (\ref{b9}) are called the Manley--Rowe--Weiss (MRW) equations \cite{man56,wei57}. The first equation states that the total photon flux is conserved, whereas the second states that signal and idler photons are created (or destroyed) in pairs ($2\pi_p \leftrightarrow \pi_r + \pi_s$, where $\pi_i$ represents a photon with frequency $\om_i$).

Suppose that the pump is strong, whereas the signal and idler (sidebands) are weak. Then one can neglect the terms in Eqs. (\ref{b3}) -- (\ref{b5}) that are of second (or third) order in the sideband amplitudes. By doing so, one obtaines the reduced equations
\ba d_z A_p &= &i(\bt_p + \ga_3|A_p|^2)A_p, \label{b11} \\
d_z A_r &= &i(\bt_r + 2\ga_3|A_p|^2)A_r + i\ga_3A_p^2A_s^*, \label{b12} \\
d_z A_s &= &i(\bt_s + 2\ga_3|A_p|^2)A_s + i\ga_3A_p^2A_r^*. \label{b13} \ea
Notice that $A_r$ is coupled to $A_s^*$ and $A_s$ is coupled to $A_r^*$.
The pump equation (\ref{b11}) has the solution
\be A_p(z) = B_p\exp[i(\bt_p + \ga_3|B_p|^2)z], \label{b14} \ee
where $B_p$ is a constant. By making substitutions of the form $A_i(z) = B_i(z)\exp[i(\bt_p + \ga_3|B_p|^2)z]$ in Eqs. (\ref{b12}) and (\ref{b13}), one obtains the modified sideband equations
\be d_z B_r = i\de_r B_r + i\ga B_s^*, \ \ d_z B_s = i\de_s B_s + i\ga B_r^*, \label{b15} \ee
where the wavenumber mismatch $\de_i = \bt_i - \bt_p + |\ga|$ and the nonlinear coupling coefficient $\ga = \ga_3B_p^2$.
Equations (\ref{b15}) have constant coefficients, but are asymmetric ($\de_r \neq \de_s$). By making the substitutions $B_r(z) = C_r(z)\exp[i(\de_r - \de_s)z/2]$ and $B_s(z) = C_s(z)\exp[i(\de_s - \de_r)z/2]$ in Eq. (\ref{b15}), one obtains the symmetrized equations
\be d_z C_r = i\de C_r + i\ga C_s^*, \ \ d_z C_s = i\de C_s + i\ga C_r^*, \label{b16} \ee
where the (common) mismatch, $\de = (\de_r + \de_s)/2 = (\bt_r + \bt_s)/2 - \bt_p + |\ga|$, depends on the average of the sideband wavenumbers.

Equations (\ref{b16}) can be written in the matrix form
\be {d \over dz} \left[\begin{array}{c} C_s \\ C_r^* \end{array}\right]
= \left[\begin{array}{cc} i\de & i\ga \\ -i\ga^* & -i\de \end{array}\right]
\left[\begin{array}{c} C_s \\ C_r^* \end{array}\right]. \label{b17} \ee
Notice that the coefficient (generating) matrix $G = iSH$, where $S =\diag(1, -1)$ is the metric matrix defined in Sec. 2.2 and $H$ is hermitian. Notice also that $G = \de G_3 + \ga_rG_2 - \ga_iG_1$, where the generators of SU(1,1) were defined in Eq. (\ref{2.2.14}). The solution of Eq. (\ref{b17}) can be written in the input--output form
\be \left[\begin{array}{c} C_s(z) \\ C_r^*(z) \end{array}\right]
= \left[\begin{array}{cc} C + i\de S/k & i\ga S/k \\ -i\ga^*S/k &C - i\de S/k \end{array}\right]
\left[\begin{array}{c} C_s(0) \\ C_r^*(0) \end{array}\right], \label{b18} \ee
where $C = \cosh(kz)$, $S = \sinh(kz)$ and $k = (|\ga|^2 - \de^2)^{1/2}$. Provided that $|\ga| > \de$, the sideband amplitudes grow with distance. Notice that the transfer matrix has the canonical form of Eq. (\ref{2.2.12}).

In nondegenerate FWM, which is also called parametric amplification \cite{mck04}, two pumps ($p$ and $q$) interact with two sidebands ($\pi_p + \pi_q \leftrightarrow \pi_r + \pi_s$). In the standard configuration, the high- and low-frequency waves are pumps, whereas the intermediate-frequency waves are sidebands ($\om_p < \om_r < \om_s < \om_q$). By following the procedure described above, one obtains sideband equations of the form (\ref{b16}), where the wavenumber mismatch $\de = (\bt_r + \bt_s - \bt_p - \bt_q)/2 + \ga_3(|B_p|^2 + |B_q|^2)/2$ and the nonlinear coupling coefficient $\ga = 2\ga_3B_pB_q$. Notice that the mismatch depends on the average of the pump wavenumbers and powers, whereas the coupling coefficient depends on the product of the pump amplitudes and has the nondegeneracy factor 2.

Light-wave propagation in a second-order nonlinear medium is governed by an equation similar to (\ref{b1}), in which the nonlinear term is $\ga_2A^2$ \cite{boy20}. In three-wave mixing (TWM), which is also called parametric down-conversion, a pump wave interacts with signal and idler waves, subject to the frequency-matching condition $\om_p = \om_r + \om_s$.
By substituting ansatz (\ref{b2}) in the wave equation and collecting terms of like frequency, one obtains the amplitude equations
\ba d_z A_p &= &i\bt_p A_p + i\ga_2A_rA_s, \label{b21} \\
d_z A_r &= &i\bt_r A_r + i\ga_2A_pA_s^*, \label{b22} \\
d_z A_s &= &i\bt_s A_s + i\ga_2A_pA_r^*. \label{b23} \ea
By combining Eqs. (\ref{b21}) -- (\ref{b23}), one obtains the MRW equations
\be d_z (|A_p|^2 + |A_r|^2) = 0, \ \ d_z (|A_r|^2 - |A_s|^2) = 0. \label{b24} \ee
Once again, signal and idler photons are created (or destroyed) in pairs ($\pi_p \leftrightarrow \pi_r + \pi_s$).
By following the procedure described above, one obtains linearized equations of the form (\ref{b16}), where the wavenumber mismatch $\de = (\bt_r + \bt_s - \bt_p)/2$ and the nonlinear coupling coefficient $\ga = \ga_2B_p$. Notice that there is no nonlinear contribution to the mismatch.

In App. A, we showed that there exists a complex Hamiltonian formalism for simple harmonic oscillators.
By combining the Hamiltonian
\be H = \de(|C_r|^2 + |C_s|^2) + \ga C_r^*C_s^* + \ga^*C_rC_s \label{b25} \ee
with the Hamilton equation
\be d_z C_i = i\pd H/\pd C_i^*, \label{b26} \ee
one obtains the signal and idler equations (\ref{b16}). This formalism is a natural bridge between the classical and quantum models of parametric amplification.

\newpage

\sec{Appendix C: Frequency conversion}

Consider light-wave propagation in a third-order nonlinear medium (App. B). There are variants of nondegenerate four-wave mixing, in which two pump waves ($p$ and $q$) interact with signal and idler waves ($s$ and $r$), subject to the frequency-matching condition $\om_p + \om_s $ $\om_q + \om_r$ \cite{mck04}. In nearby frequency conversion, $\om_p < \om_q < \om_r < \om_s$, whereas in distant frequency conversion, $\om_p < \om_r < \om_q < \om_s$.
By substituting the four-frequency ansatz
\ba A(t, z) &= &A_p(z)\exp(-i\om_pt) + A_q(z)\exp(-i\om_qt) \nonumber \\
&&+\ A_r(z)\exp(-i\om_rt) + A_s(z)\exp(-i\om_st) \label{c1} \ea
in Eq. (\ref{b1}) and collecting terms of like frequency, one obtains the amplitude equations
\ba d_z A_p &= &i(\bt_p + \ga_3|A_p|^2 + 2\ga_3|A_q|^2 + 2\ga_3|A_r|^2 + 2\ga_3|A_s|^2)A_p + i2\ga_3A_qA_rA_s^*, \label{c2} \\
d_z A_q &= &i(\bt_q + 2\ga_3|A_p|^2 + \ga_3|A_q|^2 + 2\ga_3|A_r|^2 + 2\ga_3|A_s|^2)A_q + i2\ga_3A_pA_r^*A_s, \label{c3} \\
d_z A_r &= &i(\bt_r + 2\ga_3|A_p|^2 + 2\ga_3|A_q|^2 + \ga_3|A_r|^2 + 2\ga_3|A_s|^2)A_r + i2\ga_3A_pA_q^*A_s, \label{c4} \\
d_z A_s &= &i(\bt_s + 2\ga_3|A_p|^2 + 2\ga_3|A_q|^2 + 2\ga_3|A_r|^2 + \ga_3|A_s|^2)A_s + i2\ga_3A_p^*A_qA_r, \label{c5} \ea
where $\bt_i = \bt(\om_i)$.
The factors of 1 and 2 that precede $\ga_3$ are called nondegeneracy factors.
By combining Eqs. (\ref{c2}) -- (\ref{c5}), one finds that
\ba d_z |A_p|^2 &= &i2\ga_3A_p^*A_qA_rA_s^* - i2\ga_3A_pA_q^*A_r^*A_s, \label{c6} \\
d_z |A_q|^2 &= &i2\ga_3A_pA_q^*A_r^*A_s - i2\ga_3A_p^*A_qA_rA_s^*, \label{c7} \\
d_z |A_r|^2 &= &2i\ga_3A_pA_q^*A_r^*A_s - 2i\ga_3A_p^*A_qA_rA_s^* , \label{c8} \\
d_z |A_s|^2 &= &2i\ga_3A_p^*A_qA_rA_s^* - 2i\ga_3A_pA_q^*A_r^*A_s, \label{c9} \ea
from which it follows that
\be d_z (|A_p|^2 + |A_q|^2) = 0, \ \ 
d_z (|A_r|^2 + |A_s|^2) = 0. \label{c10} \ee
Equations (\ref{c10}) are called the Manley--Rowe--Weiss (MRW) equations \cite{man56,wei57}. The first equation states that the total pump flux is conserved, whereas the second states that the total signal and idler (sideband) flux is conserved
($\pi_p + \pi_s \leftrightarrow \pi_q + \pi_r$, where $\pi_i$ represents a photon with frequency $\om_i$).
In (third-order) frequency conversion, power flows from the high- and low-frequency waves to the intermediate-frequency waves (or {\it vice versa}), just as it does in parametric amplification. The differences between the processes are which waves are strong and which waves are weak. (The roles of waves $q$ and $s$ are interchanged.)

Suppose that the pumps are strong and the sidebands are weak. Then one can neglect the terms in Eqs. (\ref{c2}) -- (\ref{c5}) that are of second (or third) order in the sideband amplitudes. By doing so, one obtains the reduced equations
\ba d_z A_p &= &i(\bt_p + \ga_3|A_p|^2 + 2\ga_3|A_q|^2)A_p, \label{c11} \\
d_z A_q &= &i(\bt_q + 2\ga_3|A_p|^2 + \ga_3|A_q|^2)A_q, \label{c12} \\
d_z A_r &= &i(\bt_r + 2\ga_3|A_p|^2 + 2\ga_3|A_q|^2)A_r + i2\ga_3A_pA_q^*A_s, \label{c13} \\
d_z A_s &= &i(\bt_s + 2\ga_3|A_p|^2 + 2\ga_3|A_q|^2)A_s + i2\ga_3A_p^*A_qA_r. \label{c14} \ea
Notice that $A_r$ is coupled to $A_s$ and $A_s$ is coupled to $A_r$.
The pump equations (\ref{c11}) and (\ref{c12}) have the solutions
\ba A_p(z) &= &B_p\exp[i(\bt_p + \ga_3|B_p|^2 + 2\ga_3|B_q|^2)z], \label{c15} \\
 A_q(z) &= &B_q\exp[i(\bt_q + 2\ga_3|B_p|^2 + \ga_3|B_q|^2)z], \label{c16} \ea
where $B_p$ and $B_q$ are constants. By making the substitutions $A_r(z) = B_r(z)\exp[i(\bt_p + \ga_3|B_p|^2 + 2\ga_3|B_q|^2)z]$ and $A_s(z) = B_s(z)\exp[i(\bt_q + 2\ga_3|B_p|^2 + \ga_3|B_q|^2)z]$ in Eqs. (\ref{c13}) and (\ref{c14}), one obtains the modified sideband equations
\be d_z B_r  = i\de_rB_r + i\ga^*B_s, \ \ d_z B_s = i\de_sB_s + i\ga B_r, \label{c17} \ee
where the wavenumber mismatches $\de_r = \bt_r - \bt_p + \ga_3|B_p|^2$ and $\de_s = \bt_s - \bt_q + \ga_3|B_q|^2$, and the nonlinear coupling coefficient $\ga = 2\ga_3B_p^*B_q$. By making the substitutions $B_i(z) = C_i(z)\exp[i(\de_r + \de_s)z/2]$ in Eqs. (\ref{c17}), one obtains the alternative equations
\be d_z C_r = -i\de C_r + i\ga^*C_s, \ \ d_z C_s = i\de C_s + i\ga C_r, \label{c18} \ee
where the mismatch, $\de = (\de_s - \de_r)/2 = (\bt_p + \bt_s - \bt_q - \bt_r)/2 + \ga_3(|B_q|^2 - |B_p|^2)/2$, depends on the differences between the pump (sideband) wavenumbers and pump powers.

Equations (\ref{c18}) can be written in the matrix form
\be {d \over dz} \left[\begin{array}{c} C_s \\ C_r \end{array}\right]
= \left[\begin{array}{cc} i\de & i\ga \\ i\ga^* & -i\de \end{array}\right]
\left[\begin{array}{c} C_s \\ C_r \end{array}\right]. \label{c19} \ee
Notice that the coefficient (generating) matrix $G = iH$, where $H$ is hermitian. Notice also that $G = \de G_1 + \ga_rG_2 + \ga_iG_3$ or, equivalently, $H = \de\si_1 + \ga_r\si_2 - \ga_i\si_3$, where the generators of SU(2) were defined in Eqs. (\ref{2.2.4}) and (\ref{2.2.6}). The solution of Eq. (\ref{c19}) can be written in the input--output form
\be \left[\begin{array}{c} C_s(z) \\ C_r(z) \end{array}\right]
= \left[\begin{array}{cc} c + i\de s/k & i\ga s/k \\ i\ga^*s/k &c - i\de s/k \end{array}\right]
\left[\begin{array}{c} C_s(0) \\ C_r(0) \end{array}\right], \label{c20} \ee
where $c = \cos(kz)$, $s = \sin(kz)$ and $k = (|\ga|^2 + \de^2)^{1/2}$. As distance increases, the signal and idler exchange power periodically. Neither sideband amplitude grows without bound (App. B). Notice that the transfer matrix has the canonical form of Eq. (\ref{2.2.2}).

Now consider light-wave propagation in a second-order nonlinear medium (App. B). 
There is a variant of three-wave mixing, in which a pump wave interacts with signal and idler waves, subject to the frequency-matching condition $\om_p + \om_r = \om_s$. By substituting the three-frequency ansatz (\ref{b2}) in the wave equation and collecting terms of like frequency, one obtains the amplitude equations
\ba d_z A_p &= &i\bt_p A_p + i\ga_2A_r^*A_s, \label{c21} \\
d_z A_r &= &i\bt_r A_r + i\ga_2A_p^*A_s, \label{c22} \\
d_z A_s &= &i\bt_s A_s + i\ga_2A_pA_r. \label{c23} \ea
By combining Eqs. (\ref{c21}) -- (\ref{c23}), one obtains the MRW equations
\be d_z (|A_p|^2 - |A_r|^2) = 0, \ \ d_z (|A_r|^2 + |A_s|^2) = 0. \label{c24} \ee
The first equation states that pump and idler photons are created (or destoyed) in pairs, whereas the second states that the total signal and idler flux is conserved ($\pi_p + \pi_r \leftrightarrow \pi_s$).
In (second-order) frequency conversion, power flows from the high-frequency wave to the lower-frequency waves (or {\it vice versa}), just as it does in parametric down-conversion. The differences between the processes are which wave is strong and which waves are weak. (The roles of waves $p$ and $s$ are interchanged.)
By following the procedure described above, one obtains linearized equations of the form (\ref{c18}), where the wavenumber mismatch $\de = (\bt_s - \bt_p - \bt_r)/2$ and the nonlinear coupling coefficient $\ga = \ga_2B_p$. Notice that there is no nonlinear contribution to the mismatch.

A complex Hamiltonian formalism also exists for frequency conversion. By combining the Hamiltonian
\be H = \de(|C_s|^2 - |C_r|^2) + \ga C_s^*C_r + \ga^*C_sC_r^* \label{c25} \ee
with the Hamilton equation
\be d_z C_i = i\pd H/\pd C_i^*, \label{c26} \ee
one obtains the signal and idler equations (\ref{c18}). This formalism is a natural bridge between the classical and quantum models of frequency conversion.

\newpage

\sec{Appendix D: Rotation}

Consider a rotation in three dimensions. Let $\vn = (n_1, n_2, n_3)$ be a unit vector parallel to the rotation axis and let $\th$ be the rotation angle. In addition, let $\vv$ and $\vw$ be the vector of interest before and after the rotation. Then the parallel component $\vv_\pa = \vn(\vn\cdot\vv)$ is not affected by the rotation. The perpendicular component $\vv_\pe = \vv - \vv_\pa = (1 - \vn\vn\cdot)\vv$. The vectors $\vn$ and $\vv_\pe$ define two perpendicular axes, and the third axis is parallel to $\vn \times \vv_\pe$. Rotation changes $\vv$ into
\ba \vw &= &\vn\vn\cdot\vv + c(1 - \vn\vn\cdot)\vv + s\vn\times(1 - \vn\vn\cdot)\vv \nonumber \\
&= &c\vv + s\vn\times\vv + (1 - c)\vn\vn\cdot\vv, \label{d1} \ea
where $c = \cos(\th)$ and $s = \sin(\th)$.
In matrix form, the two operators in Eq. (\ref{d1}) are
\be \vn \times = \left[\begin{array}{ccc} 0 & -n_3 & n_2 \\ n_3 & 0 & -n_1 \\ -n_2 & n_1 & 0 \end{array}\right], \ \ 
\vn\vn\cdot = \left[\begin{array}{ccc} n_1^2 & n_1n_2 & n_1n_3 \\ n_2n_1 & n_2^2 & n_2n_3 \\ n_3n_1 & n_3n_2 & n_3^2 \end{array}\right]. \label{d2} \ee
Notice that the first matrix in Eq. (\ref{d2}) is asymmmetric, whereas the second is symmetric.
By combining Eqs. (\ref{d1}) and (\ref{d2}), one obtains the rotation matrix
\be R = \left[\begin{array}{ccc} c + n_1^2d & -n_3s + n_1n_2d & n_2s + n_1n_3d \\ n_3s + n_2n_1d & c + n_2^2d & -n_1s + n_2n_3d \\ -n_2s + n_3n_1d & n_1s + n_3n_2d & c + n_3^2d \end{array}\right], \label{d3} \ee
where $d = 1 - c$. Equation (\ref{d3}), which specifies $R = [r_{ij}]$ in terms of $\vn$ and $\th$, was stated in Sec. 2.3. If $R$ is specified, then it follows from the diagonal terms in Eq. (\ref{d3}) that
\be c = [\tr(R) - 1]/2, \ \ s = (1 - c^2)^{1/2}, \label{d4} \ee
and it follows from the asymmetric terms that
\be n_1 = (r_{32} - r_{23})/2s, \ \ n_2 = (r_{13} - r_{31})/2s, \ \ n_3 = (r_{21} - r_{12})/2s. \label{d5} \ee
These equations can be written in the compact form $n_i = (R^t - R)_{jk}/2s$, where $i$, $j$ and $k$ are in positive cyclic order. In Eqs. (\ref{d3}) and (\ref{d5}), changing the sign of $s$ is equivalent to changing the signs of $n_i$, so one can assume that $s  > 0$ without loss of generality.

With $\vv$ and $\vw$ regarded as column vectors, Eq. (\ref{d1}) can be rewritten in the matrix form $\vw = R\vv$. If the input vector $\vv$ and rotation matrix $R$ are specified, then the output vector $\vw$ is defined by the preceding equation. However, if $\vv$ and $\vw$ are specified, then one constructs the required $R$ by defining
\be \vn = \vv \times \vw/|\vv \times \vw|, \ \ \cos(\th) = (\vv\cdot\vw)/|\vv||\vw|. \label{d6} \ee

\newpage

\sec{Appendix E: Lorentz transformation}

Consider a Lorentz transformation in time and two space dimensions, in which context the coordinate vector $X = [t, x, y]^t$, and let $L = [l_{ij}]$ be the transformation matrix. Then, as stated in Sec. 2.3, $L$ satisfies the equivalent equations
\be L^tSL = S, \ \ L^{-1} = SL^tS, \label{e1} \ee
where $S = \diag(1, -1, -1)$ is the metric matrix.
The first of Eqs. (\ref{e1}) ensures that the spacetime interval $X^tSX = t^2 - x^2 - y^2$ is conserved.
This matrix equation involves nine scalar equations for the components $l_{ij}$. But $(L^tSL)^t = L^tSL$, so only six of these equations are independent. Hence, $L$ is specified by three free parameters. Examples of Lorentz matrices include the identity matrix (which has no free parameters), and rotation and boost matrices (which have one and two free parameters, respectively). None of these examples has three free parameters, so they are special cases of Lorentz matrices.

In this appendix, we derive the general form of a Lorentz matrix. (This derivation is also provided in \cite{mck25b}.)
It is convenient to write
\be L = \left[\begin{array}{cc} \ga & R \\ C & M \end{array}\right], \label{e2} \ee
where $C$ is a $2 \times 1$ column vector, $R$ is a $1 \times 2$ row vector and $M$ is a $2 \times 2$ matrix. It follows from Eqs. (\ref{e1}) and (\ref{e2}) that
\be L^{-1} = \left[\begin{array}{cc} \ga & -C^t \\ -R^t & M^t \end{array}\right]. \label{e3} \ee
%
By combining Eqs. (\ref{e2}) and (\ref{e3}), one finds that
\ba \left[\begin{array}{cc} \ga & -C^t \\ -R^t & M^t \end{array}\right]
\left[\begin{array}{cc} \ga & R \\ C & M \end{array}\right]
&= &\left[\begin{array}{cc} \ga^2 - C^tC & \ga R - C^tM \\ M^tC - \ga R^t & M^tM - R^tR \end{array}\right], \label{e4} \\
\left[\begin{array}{cc} \ga & R \\ C & M \end{array}\right]
\left[\begin{array}{cc} \ga & -C^t \\ -R^t & M^t \end{array}\right]
&= &\left[\begin{array}{cc} \ga^2 - RR^t & RM^t - \ga C^t \\ \ga C - MR^t & MM^t - CC^t \end{array}\right]. \label{e5} \ea
The matrices on the right sides of Eqs. (\ref{e4}) and (\ref{e5}) should equal the identity matrix $I$.

It follows from the top-left entries of Eqs. (\ref{e4}) and (\ref{e5}) that
\be \ga^2 - 1 = C^tC = RR^t. \label{e6} \ee
Hence, $C$ and $R$ have the same length, $u = (\ga^2 - 1)^{1/2}$, in which case $C = u[\cos(\th_2), \sin(\th_2)]^t$ and $R =u [\cos(\th_1), \sin(\th_1)]$.
The scalar $\ga$, and the vectors $C$ and $R$, are specified by three parameters: $\ga$ itself and the angles $\th_1$ and $\th_2$. No free parameters remain, so $M$ must be specified by scalar functions of $\ga$ and matrix combinations of $C$ and $R$. The top-right entry of Eq. (\ref{e4}) requires that $C^tM = \ga R$ and the bottom-left entry of Eq. (\ref{e5}) requires that $MR^t = \ga C$. Of the matrices $CC^t$, $R^tR$, $CR$ and $R^tC^t$, only the third has the property that $C^t(CR) \propto R$ and $(CR)R^t \propto C$. Hence, we choose the ansatz $M = N + \ep CR$, where the matrix $N$ and scalar $\ep$ remain to be determined.

The bottom-left entry of Eq. (\ref{e5}) is
\be 0 = \ga C - MR^t = \ga C - (NR^t + \ep u^2 C). \label{e7} \ee
This equation requires that $NR^t = C$, so $N$ is the rotation matrix that converts $R^t$ to $C$. The subsequent equation $\ga - 1 = \ep u^2$ requires that $\ep = 1/(\ga + 1)$. With $N$ and $\ep$ so defined, the ansatz satisfies the bottom-left equation.

The top-right entry of Eq. (\ref{e4}) is
\be 0 = \ga R - C^tM = \ga R - (C^tN + \ep u^2 R). \label{e8} \ee
The identity $NR^t = C$ implies that $C^tN = RN^tN = R$, as required, and the identity $\ep u^2$ $= \ga - 1$ ensures that the top-right equation is satisfied.

The bottom-right entry of matrix (\ref{e5}) is
\ba MM^t - CC^t &= &(N + \ep CR)(N^t + \ep R^tC^t) - CC^t \nonumber \\
&= &I + \ep CC^t + \ep CC^t + \ep^2u^2 CC^t - CC^t. \label{e9} \ea
The $CC^t$ terms cancel, because $\ep(2 + \ep u^2) = 1$, so $MM^t - CC^t = I$, as required. The $R^tR$ terms in the bottom-right entry of matrix (\ref{e4}) cancel for the same reason.

The preceding results are summarized by the equations
\ba \left[\begin{array}{cc} \ga & R \\ C & N + \ep CR \end{array}\right]
&= &\left[\begin{array}{cc} 1 & 0 \\ 0 & N \end{array}\right]
\left[\begin{array}{cc} \ga & R \\ R^t & I + \ep R^tR \end{array}\right] \nonumber \\
&= &\left[\begin{array}{cc} \ga & C^t \\ C & I + \ep CC^t \end{array}\right]
\left[\begin{array}{cc} 1 & 0 \\ 0 & N \end{array}\right]. \label{e10} \ea
Thus, a Lorentz transformation consists of a boost followed by a rotation [Eqs. (\ref{4.2.5}) and (\ref{4.2.6})], or a rotation followed by a different boost. The free parameters are the energy ($\ga$) and direction angle of the boost ($\th_1$ or $\th_2$), and the rotation angle ($\th_{21} = \th_2 - \th_1$). Written explicitly,
\ba \left[\begin{array}{cc} \ga & R \\ C & N + \ep CR \end{array}\right]
&= &\left[\begin{array}{ccc} \ga & uc_1 & us_1 \\ uc_2 & c_{21} + \ep(uc_2)(uc_1) & -s_{21} + \ep(uc_2)(us_1) \\
us_2 & s_{21} + \ep(us_2)(uc_1) & c_{21} + \ep(us_2)(us_1) \end{array}\right] \nonumber \\
&= &\left[\begin{array}{ccc} \ga & uc_1 & us_1 \\ uc_2 & c_{21} + \de c_2c_1 & -s_{21} + \de c_2s_1) \\
us_2 & s_{21} + \de s_2c_1 & c_{21} + \de s_2s_1 \end{array}\right], \label{e11} \ea
where $c_i = \cos(\th_i)$, $c_{21} = \cos(\th_{21})$ and $\de = \ep u^2 = \ga - 1$. The definitions of $s_i$ and $s_{21}$ are similar. Equation (\ref{e11}), which specifies $L$ in terms of $\ga$, $\th_1$ and $\th_2$, was stated in Sec. 2.3. If $L$ is specified, then $\ga = l_{11}$, $\tan(\th_1) = l_{13}/l_{12}$ and $\tan(\th_2) = l_{31}/l_{21}$.

\newpage

\sec{Appendix F: Jones--Stokes formalism}

In this appendix, we propose a Jones--Stokes formalism for SU(1,1) and SO(1,2).
For every (complex) Jones vector $|s\> = [u, v]^t$, there exists an associated (real) Stokes-like vector $\vs = [t, x, y]^t = [s_1, s_2, s_3]^t$. The Stokes components are defined by the equations
\be s_1 = |u|^2 + |v|^2, \ \ s_2 = u^*v + v^*u, \ \ s_3 = i(u^*v - v^*u). \label{f1} \ee
SU(1,1) matrix operations in Jones space preserve the norm $|u|^2 - |v|^2$, whereas SO(1,2) operations in Stokes space preserve the norm $s_1^2 - s_2^2 - s_3^2 = (|u|^2 - |v|^2)^2$. Definitions (\ref{f1}) differ from definitions (\ref{6.3.31}) -- (\ref{6.3.33}) in two ways: First, $s_1$ is the sum of $|u|^2$ and $|v|^2$, rather than the difference, and $s_3$ includes the factor $i$, rather than $1/i$. The first change ensures that the Stokes norm equals $(|u|^2 - |v|^2)^2$, whereas the second ensures that the Stokes transformations are active.

The (reordered) generators of SU(1,1) are
\be G_1 = \left[\begin{array}{cc} i & 0 \\ 0 & -i \end{array}\right], \ \ 
G_2 = \left[\begin{array}{cc} 0 & i \\ -i & 0 \end{array}\right], \ \ 
G_3 = \left[\begin{array}{cc} 0 & 1 \\ 1 & 0 \end{array}\right], \label{f2} \ee
and the associated fundamental matrices $M_i = \exp(G_ik_i)$ are
\be M_1 = \left[\begin{array}{cc} e_1 & 0 \\ 0 & e_1^* \end{array}\right], \ \ 
M_2 = \left[\begin{array}{cc} C_2 & iS_2 \\ -iS_2 & C_2 \end{array}\right], \ \ 
M_3 = \left[\begin{array}{cc} C_3 & S_3 \\ S_3 & C_3 \end{array}\right]. \label{f3} \ee

Under transformation 1, $u' = eu = (c + is)u$ and $v' = e^*v = (c - is)v$, from which it follows that
\ba |u'|^2 + |v'|^2 &= &|u|^2 + |v|^2, \\
u'^*v' + v'^*u' &= &u^*v(d - it) + v^*u(d + it) \nonumber \\
&= &d(u^*v + v^*u) - ti(u^*v - v^*u), \\
i(u'^*v' - v'^*u') &= &i[u^*v(d - it) - v^*u(d + it)] \nonumber \\
&= &t(u^*v + v^*u) + di(u^*v - v^*u), \label{f4} \ea
where $d = c^2 - s^2$, $t = 2cs$ and the subscripts 1 were omitted.

Under transformation 2, $u' = Cu + iSv$ and $v' = Cv - iSu$, from which it follows that%
\ba |u'|^2 + |v'|^2 &= &(Cu^* - iSv^*)(Cu + iSv) + (Cv^* + iSu^*)(Cv - iSu) \nonumber \\
&= &C^2|u|^2 + iCS(u^*v - v^*u) + S^2|v|^2 \nonumber \\
&&+\ C^2|v|^2 + iCS(u^*v - v^*u) + S^2|u|^2 \nonumber \\
&= &D(|u|^2 + |v|^2) + Ti(u^*v - v^*u), \\
u'^*v' + v'^*u' &= &(Cu^* - iSv^*)(Cv - iSu) + (Cv^* + iSu^*)(Cu + iSv) \nonumber \\
&= &C^2u^*v - iCS(|u|^2 + |v|^2) - S^2v^*u \nonumber \\
&&+\ C^2v^*u +iCS(|u|^2 + |v|^2) - S^2u^*v \nonumber \\
&= &u^*v + v^*u, \\
i(u'^*v' - v'^*u') &= &i\{C^2u^*v - iCS(|u|^2 + |v|^2) - S^2v^*u \nonumber \\
&&-\ [C^2v^*u +iCS(|u|^2 + |v|^2) - S^2u^*v]\} \nonumber \\
&= &T(|u|^2 + |v|^2) + Di(u^*v - v^*u), \label{f5} \ea
where $D = C^2 + S^2$ (not $C -1$) and $T = 2CS$.

Under transformation 3, $u' = Cu + Sv$ and $v' = Cv + Su$, from which it follows that%
\ba |u'|^2 + |v'|^2 &= &(Cu^* + Sv^*)(Cu + Sv) + (Cv^* + Su^*)(Cv + Su) \nonumber \\
&= &C^2|u|^2 + CS(u^*v + v^*u) + S^2|v|^2 \nonumber \\
&&+\ C^2|v|^2 + CS(u^*v + v^*u) + S^2|u|^2 \\
&= &D(|u|^2 + |v|^2) + T(u^*v + v^*u), \\
u'^*v' + v'^*u' &= &(Cu^* + Sv^*)(Cv + Su) + (Cv^* + Su^*)(Cu + Sv) \nonumber \\
&= &C^2u^*v + S^2v^*u + CS(|u|^2 + |v|^2) \nonumber \\
&&+\ C^2v^*u + S^2u^*v + CS(|u|^2 + |v|^2) \nonumber \\
&= &T(|u|^2 + |v|^2) + D(u^*v + v^*u), \\
i(u'^*v' - v'^*u') &= &i\{C^2u^*v + S^2v^*u + CS(|u|^2 + |v|^2) \nonumber \\
&&-\ [C^2v^*u + S^2u^*v + CS(|u|^2 + |v|^2)]\} \nonumber \\
&= &i(u^*v - v^*u). \label{f6} \ea
It is easy to verify that all three transformations preserve $|u|^2 - |v|^2$, as stated above.

By rewriting the preceding results in matrix form, one obtains the associated
Lorentz matrices
\be L_1 = \left[\begin{array}{ccc} 1 & 0 & 0 \\ 0 & d & -t \\ 0 & t & d \end{array}\right], \ \ 
L_2 = \left[\begin{array}{ccc} D & 0 & T \\ 0 & 1 & 0 \\ T & 0 & D \end{array}\right], \ \ 
L_3 = \left[\begin{array}{ccc} D & T & 0 \\ T & D & 0 \\ 0 & 0 & 1 \end{array}\right]. \label{f7} \ee
Each operation $L_i$ preserves the Stokes component $s_i$. The first operation produces a rotation in the $xy$ plane, the second produces a Lorentz boost in the $yt$ plane and the third produces a boost in the $tx$ plane. All three transformations are active. If we had defined $s_3$ with the factor $1/i$, the transformations would have been passive.

There is another rationale for changing the sign of $s_3$. In \cite{mck13}, it was stated that for SU(1,1), the generating matrix $G = iSH$, where $S = \diag(1, -1)$ is the metric matrix and
\ba H &= &\left[\begin{array}{cc} \de & (\ga_r + i\ga_r) \\ (\ga_r - i\ga_i) & \de \end{array}\right] \nonumber \\
&= &\de \left[\begin{array}{cc} 1 & 0 \\ 0 & 1 \end{array}\right]
+ \ga_r \left[\begin{array}{cc} 0 & 1 \\ 1 & 0 \end{array}\right]
+ \ga_i\left[\begin{array}{cc} 0 & i \\ -i & 0 \end{array}\right] \label{f8} \ea
is a Hermitian matrix. In the context of parametric amplification (App. C), $\de$ is the wave-number mismatch coefficient and $\ga$ is the nonlinear coupling coefficient. If one uses the matrices in Eq. (\ref{f8}), which are denoted by $H_1$, $H_2$ and $H_3$, respectively, to define the Stokes components  $s_i = \<s|H_i|s\>$, one finds that
\ba s_1 &= &[u^*, v^*][u, v]^t \ = \ |u|^2 + |v|^2, \label{f9} \\
s_2 &= &[u^*, v^*][v, u]^t \ = \ u^*v + v^*u, \label{f10} \\
s_3 &= &[u^*, v^*][iv, -iu]^t \ = \ i(u^*v - v^*u). \label{f11} \ea
In this approach, $s_3$ has the opposite sign naturally.
It is easy to verify that
\ba &&M_1^\d H_1M_1 = H_1, \ \ M_1^\d H_2M_1 = dH_2 - tH_3, \ \ M_1^tH_3M_1 = dH_3 + tH_2, \label{f12} \\
&&M_2^\d H_1M_2 = DH_1 + TH_3, \ \ M_2^\d H_2M_2 = H_2, \ \ M_2^tH_3M_2 = DH_3 + TH_1, \label{f13} \\
&&M_3^\d H_1M_3 = DH_1 + TH_2, \ \ M_3^\d H_2M_3 = DH_2 + TH_1, \ \ M_3^tH_3M_3 = H_3. \label{f14} \ea
By defining the vector $\vh = [H_1, H_2, H_3]^t$  and writing the preceding results in the matrix form $\vh' = L\vh$, one obtains the Lorentz matrices (\ref{f7}).

\end{document}